\documentclass[12pt]{article}

\pdfoutput=1

\usepackage{graphics}
\usepackage{epsfig}
\usepackage{amsfonts}
\usepackage{amssymb}
\usepackage{amsmath}
\usepackage{dsfont}
\usepackage{pifont}
\usepackage{bbm}
\usepackage{multirow}
\usepackage{latexsym}
\usepackage{verbatim}
\usepackage{mcite}
\usepackage{cite}
\usepackage{units}
\usepackage[all]{xy}
\usepackage{cancel}
\usepackage{slashed}

\def\bfone{\relax{\rm 1\kern-.35em 1}}

\newcommand{\be}{\begin{equation}}
\newcommand{\ee}{\end{equation}}
\newcommand{\ben}{\begin{displaymath}}
\newcommand{\een}{\end{displaymath}}
\newcommand{\bea}{\begin{eqnarray}}
\newcommand{\eea}{\end{eqnarray}}

\newcommand{\bean}{\begin{eqnarray*}}
\newcommand{\eean}{\end{eqnarray*}}

\DeclareMathAlphabet{\mathpzc}{OT1}{pzc}{m}{it}

\begin{document}
\pagestyle{plain}


\makeatletter \@addtoreset{equation}{section} \makeatother
\renewcommand{\thesection}{\arabic{section}}
\renewcommand{\theequation}{\thesection.\arabic{equation}}
\renewcommand{\thefootnote}{\arabic{footnote}}


\setcounter{page}{1} \setcounter{footnote}{0}


\begin{titlepage}

\begin{flushright}
UUITP-11/16\\
\end{flushright}

\bigskip

\begin{center}

\vskip 0cm

{\LARGE \bf Universal isolation in the AdS landscape} \\[6mm]

\vskip 0.5cm

{\bf U.~H.~Danielsson, G.~Dibitetto  \,and\, S.~C.~Vargas}\let\thefootnote\relax\footnote{{\tt \{ulf.danielsson, giuseppe.dibitetto, sergio.vargas\} @physics.uu.se}}\\

\vskip 25pt

{\em Institutionen f\"or fysik och astronomi, University of Uppsala, \\ Box 803, SE-751 08 Uppsala, Sweden \\}

\vskip 0.8cm

\end{center}

\vskip 1cm

\begin{center}

{\bf ABSTRACT}\\[3ex]

\begin{minipage}{13cm}
\small

We study the universal conditions for quantum non-perturbative stability against bubble nucleation for pertubatively stable AdS vacua based on 
positive energy theorems. We also compare our analysis with the pre-existing ones in the literature carried out within the thin-wall approximation.
The aforementioned criterion is then tested in two explicit examples describing massive type IIA string theory compactified on $S^3$ and 
$S^3\,\times\,S^3$, respectively. The AdS landscape of both classes of compactifications is known to consist of a set of isolated points.  
The main result is that all critical points respecting the Breitenlohner-Freedaman (BF) bound also turn out be stable at a non-perturbative level.
Finally, we speculate on the possible universal features that may be extracted from the above specific examples.

\end{minipage}

\end{center}

\vfill

\end{titlepage}


\tableofcontents

\section{Introduction}
\label{sec:introduction}

In the last two decades the problem of finding suitable string compactifications giving rise to interesting lower-dimensional physics has been extensively studied adopting many different approaches.
In particular, the approach which goes under the name of \emph{flux compactification} has proven to be very effective when it comes to constructing lower-dimensional maximally symmetric vacua out of 
string theory, thus achieving complete moduli stabilization (see \emph{e.g.} \cite{Blumenhagen:2003vr,DeWolfe:2005uu}). 
The idea behind the constructions is that of pertubatively inducing a dynamical scalar potential \cite{Gukov:1999ya} for the moduli fields through the use of fluxes and branes threading
the internal manifold.

A preliminary statistical analysis of string vacua based on a counting of possible quantized flux backgrounds lead to the expectation of finding a very large amount of such maximally symmetric solutions,
which was then referred to as the string landscape \cite{Denef:2004ze}, though such an expectation is nowadays widely believed to be too na\"ive. 
Focusing in particular on vacua with negative cosmological constant (\emph{i.e.} AdS), there are hints that the AdS landscape of a given class of string compactifications might consist of
a set of \emph{isolated points}. As we will argue in this paper, there are further indications that these isolated points are non-perturbatively disconnected from each other, and that tunneling does not occur.

Generically, (warped) lower-dimensional supersymmetric AdS vacua may be obtained in string theory as the near-horizon geometry of several BPS
brane intersections \cite{Cvetic:2000cj}. Moreover, the theories of gravitational fluctuations around the supersymmetric vacuum are usually captured by an effective lower-dimensional gauged supergravity theory which may be constructed out of a
warped KK truncation \emph{Ansatz} \cite{Cvetic:2000dm,Cvetic:2000yp} that consistently relates the higher-dimensional equations of motions to simpler and algebraic lower-dimensional field equations, at
 least in the case of maximally symmetric vacua. Therefore, solving the equations of motion for the scalar fields within the effective gauged supergravity theory turns out to be a doable task,
especially by applying the techniques introduced in \cite{Dibitetto:2011gm}, where the AdS landscape of a certain class of massive type IIA compactifications was exhaustively scanned and a few novel 
additional non-supersymmetric critical points were found. One would then naturally expect to find a similar situation by applying the aforementioned techniques to different set-up's, \emph{i.e.} a 
supersymmetric AdS vacuum surrounded by a discrete set of additional critical points where internal bosonic symmetry and supersymmetry are partially or even completely broken.

Assuming the above situation to be generically realistic, and after checking perturbative stability of each point in the obtained landscape, the possibility of quantum gravitational tunneling between
different AdS vacua becomes then a very natural issue to be addressed. The aim of this paper is that of providing general criteria to assess non-perturbative stability of supergravity AdS critical points, 
 which would automatically rule out tunneling to any other point in the landscape. Our analysis will rely on the possibility of formulating positive energy theorems in AdS based on a fake-supersymmetric 
formalism. In some cases the absence of tunneling between two AdS vacua is guaranteed through the existence of an interpolating static domain wall.    

Another parallell argument in favor of an isolated AdS landscape, which may be considered rather compelling, has a holographic origin. Thanks to the AdS/CFT correspondence \cite{Maldacena:1997re}, one 
can rephrase the problem of ``charting'' the landscape of AdS string vacua into that of classifying CFT's in various dimensions. By means of the \emph{conformal bootstrap} approach 
\cite{Migdal:1972tk}, one can in principle restrict all possible conformal fixed points upon imposing conformal symmetry and self-consistency at a fundamental level. 
Great progress has recently been made in this context by using efficient numerical methods (see \emph{e.g.} \cite{Kos:2016ysd}).  
This numerical approach has so far provided strong evidence for the existence of quite a few different types of CFT's, all associated with fixed points of some RG flows. This fact singles out the role of 
conformal symmetry as an organizing principle that dictates the physical behavior of a given system in a quantum critical regime, irrespective of its microscopic description.
This suggests the emergence of universality classes of conformal fixed points.   

The above argument may be regarded as a holographic evidence for the existence of limited number of isolated and universal AdS vacua in the string landscape. However, in order for such a holographic picture to actually 
make sense, one must make sure that the corresponding gravitational vacua are stable against non-perturbative effects. In \cite{Banks:2002nm} it was in fact conjectured that a sensible theory of quantum gravity should forbid tunneling between AdS vacua. An addtional argument against tunneling provided in \cite{Banks:2002nm}, was that such a hypothetical process ending up in AdS, always leads to a big crunch. Interestingly, static domain wall solutions which interpolate between pairs of points in the landscape and hence forbidding gravitational tunneling, provide a geometrization of the corresponding RG 
flows within the dual field theory.  Therefore, the presence of a conformal fixed point in itself may be viewed as the holographic proof of the non-perturbative stability of its dual AdS vacuum.

The paper is organized as follows. In section~\ref{sec:BT}, we review the literature on bubble nucleation and the classical derivations of the corresponding 
bound on the tension of the wall separating two vacua in the thin-wall limit. 
In section~\ref{sec:DW}, we adopt a different angle on the problem of quantum non-pertubative stability in AdS. Our logic will be based on the possibility of
formulating positive energy theorems in the spirit of \cite{Townsend:1984iu,Boucher:1984yx}, which are intimately connected with fake supersymmetry. This approach
makes crucial use of the Hamilton-Jacobi (HJ) formulation of problems described by second-order dynamical equations. This will allow us to discuss
a set of universal situations that may in principle occur in the AdS landscape, some of which lead to gravitational tunneling.
In section~\ref{sec:examples}, we study in detail two explicit examples representing effective descriptions of massive type IIA string compactifications in order
to assess to what extent all of the aforementioned possiblities can be actually realized within the landscape of a consistent quantum gravity theory.
In section~\ref{sec:conclusions}, we present some concluding remarks where we try to speculate on the general features of our analysis and how to possibly extend it
to new examples. Finally, some technical details concerning the HJ formalism and pertubative techniques to solve non-linear PDE's are collected in
 appendices~\ref{appendix:HJ_flows} \& \ref{appendix:solving_PDEs}, respectively.

\section{Bubble nucleation and gravitational tunneling}
\label{sec:BT}

Before attacking the problem of quantum non-perturbative stability within a theory of gravity, we would like to start by reviewing some well-known
results concerning the mechanism of bubble nucleation.
To this end, let us consider the formation of a finite size bubble of AdS space with vacuum energy $\Lambda_2$ within another AdS space with vacuum energy 
$\Lambda_1 \, > \, \Lambda_2$. This situation was originally considered in \cite{Coleman:1980aw,Brown:1988kg}, where the conditions were derived that
would allow such a bubble to expand once spontaneously formed. When this process occurs, the true vacuum bubble would then expand and
eventually eat up all spacetime, thus completing the quantum gravitational transition.

The derivation of the condition that allows bubble nucleation works through the comparison between the total energy of the system with or without the
bubble, by properly taking into account the difference in vacuum energy as well as the positive contribution associated with the tension of the bubble
 wall. Such an energy starts out positive for a small bubble where the tension dominates and decreases as the bubble becomes larger. 
Without considering the effect of gravity, the total energy of the bubble becomes negative for sufficiently large radius, thus enabeling tunneling to
 the lower value of the vacuum energy. When gravity comes into in the game, this is no longer necessarily the case, and the energy difference may remain
 positive for any finite value of the radius leading to non-perturbative stability of the original vacuum against bubble formation. One concludes that
 gravity has a tendency to stabilize AdS vacua. Important examples are supersymmetric vacua in supergravity theories, which are protected in this way 
\cite{Witten:1981gj,Witten:1982df,Weinberg:1982id}.

Let us see how this works out in more detail by reviewing the limit of thin bubble walls. This approximation consists in neglecting variations in the 
warp factor inside the metric as one moves across the wall. This effectively implies assuming constant vacuum energy on the two sides without dynamically 
 sourcing it through the use of scalar fields. Across the wall there is a junction condition that relates the difference in the spatial curvature of the wall to the wall tension $\sigma$. We start out in static coordinates, where the metric away from the wall is then given by
\begin{equation}
ds_{4}^2= -\left( 1-\frac{\Lambda_i}{3} \rho^2 \right) dt^2 \,+\,\frac{d\rho^2}{1-\dfrac{\Lambda_i}{3} \rho^2}\,+\,\rho^2 d\Omega_{(2)}^2 \ .
\end{equation}
In case of a bubble with an inside and an outside we find
\begin{equation}
\label{junction_condition}
 \sqrt{\rho_0^{-2}-\frac{\Lambda_2}{3}}\,-\, \sqrt{\rho_0^{-2}-\frac{\Lambda_1}{3}}\,=\,\frac{\kappa_{4}^2}{2}\,\sigma \ ,
\end{equation}
where $\rho_0$ is the radius of the bubble, and we have put $\kappa_{4}^2\,=\,8 \pi G$. In order for a solution to exist for a finite value of the radius, the tension needs to obey
\begin{equation}
\label{CDL_bound}
\sigma \, \leq \,\frac{2}{\kappa_{4}^2}\, \left(\sqrt{-\frac{\Lambda_2}{3}}\,-\, \sqrt{-\frac{\Lambda_1}{3}}\right) \ ,
\end{equation}
which is commonly known as the Coleman-De Luccia (CDL) bound.
Saturation implies the existence of a bubble of infinite radius, \emph{i.e.} a straight wall. In this situation the instanton action is infinite and one does no longer have tunneling. Instead, it corresponds to a stable and static domain wall (DW) separating the two different vacua.

The same results can also be obtained though a straigtforward evaluation of the total energy. In the thin wall approximation, we find
\begin{equation}
\label{Energy_conservation}
\frac{4}{3}\,\pi\,\left( \Lambda_1\,-\, \Lambda _2 \right)\,\rho_0^{3} \,=\,
2\pi\, \kappa_{4}^2\,\sigma \,\left( \sqrt{\rho_0^{-2}-\frac{\Lambda_2}{3}}\,+\, \sqrt{\rho_0^{-2}-\frac{\Lambda_1}{3}}\right) ,
\end{equation}
where the energy of the thin wall is calculated using the average of the metric on the two sides of the wall. It is easy to check that the two equations \eqref{junction_condition} \& \eqref{Energy_conservation} have the same solution for, \emph{e.g.}, the radius $\rho_0$ as a function of the tension.

To fully capture the time evolution of a bubble it is convenient to choose a metric of the form \eqref{bubble_metric_4D}
\be
ds_{4}^{2} \ = \ e^{2a(\zeta)}\, \left[d\zeta^{2}\,- \, dt^{2} \,+\,S(t)^{2}\,\left(\frac{dr^{2}}{1\,-\,\kappa\,r^{2}} \, + \, r^{2} \,d\varphi^{2}\right)\right] \ .
\ee
The coordinate system is assumed to be comoving so that the center of the wall is always
positioned at $\zeta=0$. The Einstein equations, as well as the equations of motion in the presence of scalar fields, are discussed in detail in
 appendix~\ref{appendix:HJ_flows}. In case of pure AdS with a constant vacuum energy and without a wall, the functions $a(\zeta)$ and $S(t)$ are given by
\cite{Cvetic:1992jk}
\be
\left\{
\begin{array}{lclc} 
e^{2a(\zeta)} & = & \dfrac{\beta^2}{\alpha^2 \,\sinh^2 \left(\beta (\zeta\,-\,\zeta')\right)} & , \\[3mm]
S(t) & = & \beta^{-2} \,\cosh^2 \left(\beta \,t\right) & ,
\end{array}\right.
\ee
where $\Lambda = -\frac{3}{\alpha ^2 }$, and $\zeta'$ is chosen such that $e^{2a(0)}\,=\,1$ at the position of the wall. One can verify that these expressions can be obtained from the metric in static coordinates through a coordinate transformation. The bubble has its smallest radius, given by $1/\beta$, at $t=0$ when it appears after tunneling, and expands thereafter. 

Among the field equations, it is the first one in \eqref{DW_2nd_order} that is of particular importance, 
\begin{equation}
 3 \,  a'^2 \, - \, 3 \, q_0 \, = \, \frac{1}{2} \, K_{I J} \, \phi'^{I} \, \phi'^{J} \, - \, e^{2a} \, V \ ,
\end{equation}
from which one can recover the junction condition in case of a thin wall separating regions with different vacuum energies. Away from the bubble wall, where the system sits at a critical point of the potential with $\phi'=0$, this simply becomes
\begin{equation}
 a'^2 \, = \, q_0 \,- \,\frac{1}{3} \,e^{2a} \, V \ .
\end{equation}
Since $a'(0)$ is the extrinsic curvature of a thin wall at $\zeta=0$, we recover the junction condition, where we identify $q_0\,=\,\beta^2$ with 
$1/\rho_0^2$ at the minimum radius of the bubble.  

One can distinguish between three different kinds of bubble walls: non-extremal, extremal and ultra-extremal. The extremal case corresponds to a straight DW solution, while the ultra-extremal has a lower tension allowing for a solution with a finite radius of the  bubble. It is only bubbles with ultra-extremal walls that can form through tunneling. One may also consider non-extremal walls, which instead have a higher value of the tension than the extremal ones. These bubbles have two insides, thus resulting in a junction condition where the two spatial curvatures are {\it added} rather than subtracted. Such bubbles correspond to local maxima of the Euclidean action rather than minima and hence they cannot arise through tunneling.
The relevant summary concerning the above different types of walls can be found in table~\ref{table:Walls}.
\begin{table}[h!]
\begin{center}
\begin{tabular}{| c | c |}
\hline
Wall Type & Tension $\sigma$ \\[1mm]
\hline \hline
non-extremal & $\dfrac{2}{\kappa_{4}^2}\,\left(\sqrt{q_{0}-\dfrac{\Lambda_2}{3}}\,+\, \sqrt{q_{0}-\dfrac{\Lambda_1}{3}}\right)$ \\[1mm]
\hline
ultra-extremal & $\dfrac{2}{\kappa_{4}^2}\,\left(\sqrt{q_{0}-\dfrac{\Lambda_2}{3}}\,-\, \sqrt{q_{0}-\dfrac{\Lambda_1}{3}}\right)$\\[1mm]
\hline
extremal & $\dfrac{2}{\kappa_{4}^2}\,\left(\sqrt{-\dfrac{\Lambda_2}{3}}\,-\, \sqrt{-\dfrac{\Lambda_1}{3}}\right)$ \\[1mm]
\hline
\end{tabular}
\end{center}
\caption{{\it The different types of bubble walls according to the classification in \protect\cite{Cvetic:1992jk}. The positive parameter
$q_0$ is related to the radius of the bubble through $q_0\,=\,\rho_0^{-2}$.} 
\label{table:Walls}}
\end{table}

So far the tension of the wall has been considered as a free parameter. What we actually want to do is to form the bubble walls using the scalar fields and explicitly compute the tension of the walls. In this way, we will be able to deduce whether or not a particular vacuum is non-perturbatively stable. To proceed we need to consider the equations of motion more carefully. Most of what we need is captured by the Hamilton-Jacobi (HJ) equation determining the Hamilton principal function $F$ (see appendix~\ref{appendix:HJ_flows}),
\begin{eqnarray}
\frac{1}{12} \ e^{-2a} \ \left( \frac{\partial F}{\partial a} \right)^{2} \ - \ \frac{1}{2} \ e^{-2a} \ K^{I J} \ \frac{\partial F}{\partial \phi^I} \ \frac{\partial F}{\partial \phi^J} \ - \ 3 \ q_0 \ e^{2a} \ + \ e^{4a} \ V \ =\ 0 \ , 
\end{eqnarray}
where $\frac{\partial F }{\partial a}$ and $\frac{\partial F }{\partial \phi^I}$ appear in the first-order HJ flow equations given in \eqref{HJ_q0} 
and represent the conjugate momenta to $a$ \& $\phi^I$, respectively.
The challenge is now that of solving the above PDE to determine the function $F(a, \phi )$ satisfying the appropriate boundary conditions. 

We first consider the case of straigth walls, where $q_0=0$, in which case the equation is separable with solutions of the form \eqref{separable_F}.
As a consequence, the HJ equation becomes
\begin{equation}
V \, = \, - \, 3 \, f^2  \, + \, 2 \, K^{I J} \, \frac{\partial f}{\partial \phi^I} \, \frac{\partial f}{\partial \phi^J} \ , 
\end{equation}
which determines a function of the scalar fields that plays the role of a fake superpotential.

Let us now assume that a function $f(\phi )$ such that $f'(\phi _1)=f'(\phi _2)=0$ can be found, where $\phi _1$ and $\phi _2$ denote the positions of
 the two AdS vacua in the scalar manifold. If this is the case we can then go ahead and construct an extremal DW separating the two vacua by simply
integrating the first-order flow equations \eqref{HJ_q0=0} associated with our fake superpotential $f$. Whether such a solution actually exists is a highly non-trivial question that will be discussed at length in the next section. 

For a thin wall, across which $a$ is continous but $\phi$ and $f$ jump, we find that
the junction condition for $q_0$ is exactly saturated for a bubble of infinite size with 
\be
\begin{array}{lclc}
\Lambda _1 \ = \ -3\,f_1^2 & , & \Lambda _2 = -3\,f_2^2 & ,
\end{array}
\ee
and the tension of wall given by $\sigma \,=\, \frac{2}{\kappa_4^2}\left( f_2\,-\,f_1 \right)$, which reproduces the correct value for extremal walls
 appearing in table~\ref{table:Walls}. 

Let us now move to the case of non-vanishing $q_0$, where the HJ problem is truly non-separable. Still, the situation turns out to get considerably simplified in the thin wall approximation. Outside of the wall, one has
\begin{equation}
F(a,\phi _i )\,=\, \frac{2 \, e^{3a}}{f_{i}^{2}}\,  \left(f_i^2 \,+\,q_0 \, e^{-2a}\right)^{\frac{3}{2}} \ .
\end{equation}
Across the thin wall, where $a=0$, we find a jump in $F$, with a tension given by
\begin{equation}
\sigma \, = \, \frac{1}{3\,\kappa_4^2} \, \left(e^{-3a} \left.\frac{ \partial F}{\partial a}(a,\phi_{2}) \right|_{a=0}\,-\,e^{-3a} \left. \frac{\partial F}{\partial a}(a,\phi_{1})\right|_{a=0}\right)\, =  \, 
\frac{2}{\kappa_4^2} \, \left(\sqrt{f_2^2 + q_0}\,-\,\sqrt{f_1^2 + q_0}\right) \ .
\end{equation}
Again we find saturation, but this time for a bubble of finite size. As we will see, the crucial issue ensuring saturation is whether a function $F$
 actually exists or not for a particular value of $q_0$.

The existence of DW's is intimately connected with the property of the corresponding instantons. Instantons are given by Euclidean solutions of the form
\begin{eqnarray}
\label{round_4D_bubble}
ds_4^2 \,=\, e^{2 a(\tau)} \,\left( d \tau^2 \,+\, q_0^{-1}\, d\Omega_{(3)}^{2} \right) \ ,
\end{eqnarray}
where $\tau$ denotes Euclidean time.
The equations of motion for the instanton turn out to be identical to those for the DW with $\zeta$ replaced by $\tau$. The instanton is a 
four-dimensional bubble separating vacua at two different critical points of a given potential. Evaluating the instanton action, and finding the extrema with respect to the bubble radius, provides the junction condition or, equivalently, the expression for energy conservation. The instanton action has a finite value, corresponding to a finite probability of tunneling, only if $q_0>0$. This corresponds to a bubble forming at a finite radius after which it expands. The absence of solutions describing $\textrm{O}(4)$-symmetric bubbles of the form \eqref{round_4D_bubble} was proven  in \cite{Cvetic:1992st}
for supersymmetric AdS vacua.

\section{Fake supersymmetry \& positive energy theorems}
\label{sec:DW}
 
In the previous section we have been reviewing the general conditions for gravitational tunneling through bubble nucleation in the thin wall
 approximation. The scope of this section is that of formulating a more general sufficient criterion for non-perturbative stability in AdS which goes
beyond such an approximation. The analysis will make use of the HJ formalism presented in appendix~\ref{appendix:HJ_flows} and its relation to fake
supersymmetry.

\subsection*{A positive energy theorem}

Let us consider a theory of Einstein gravity coupled to a set of scalar fields described by an action of the form of \eqref{Full_4D_action}. Let us,
furthermore, assume that the scalar potential $V$ admits a number of perturbatively stable AdS critical points, \emph{i.e.} satisfying the
Breitenlohner-Freedman (BF) bound \cite{Breitenlohner:1982jf}. The non-perturbative stability of each AdS extremum $\phi_0$ against tunneling towards
 any of the other points in the landscape becomes a very natural issue to address. In \cite{Townsend:1984iu,Boucher:1984yx} a positive energy theorem was developed
that relies on the existence of a global function bounding the potential from below. This criterion for non-perturbative stability essentially 
generalizes the argument of \cite{Witten:1981gj} for a supersymmetric vacuum to cases which in principle have nothing to do with supersymmetry. We propose the following theorem: 

\begin{itemize}
\item {\bf Theorem:} If the scalar potential $V$ can be written as
\be\label{Vfromf_D}
V(\phi) \ = \ -\frac{D-1}{2\,(D-2)}\,f(\phi)^{2} \, + \, \frac{1}{2}\,K^{IJ}\,\partial_{I}f\,\partial_{J}f \ ,
\ee
for a suitable and \emph{globally} defined function such that
\be
\begin{array}{clc}
(i) & \partial_{I}f|_{\phi_0} \ = \ 0 & , \\[2mm]
(ii) & V(\phi) \ \geq \ -\dfrac{D-1}{2\,(D-2)}\,f(\phi)^{2} \ , \quad \forall \ \phi \, \in \, \mathcal{M}_{\textrm{scalar}} & ,
\end{array} \nonumber
\ee
then any other point in $\mathcal{M}_{\textrm{scalar}}$ has higher energy than $\phi_0$ itself and hence $\phi_0$ is stable against non-perturbative
decay. 
\end{itemize} 
Verifying the above criterion essentially boils down to the search of fake superpotentials w.r.t. which the AdS extremum in question appears to be 
``supersymmetric''. Let us now see how to use this machinery in practice.

From now on in this section, we will restrict ourselves to the case of one single scalar for the sake of simplicity, though a parallel analysis may be
carried out in cases where the scalar manifold $\mathcal{M}_{\textrm{scalar}}$ is higher-dimensional. In fact, one of these cases will be explicitly
studied in the next section. For the same reason, we will drop the factors inside \eqref{Vfromf_D} and focus on the conceptual core of the problem.

If we want to apply the above positive energy theorem at a given metastable AdS critical point $\phi_0$, we need to discuss local
and global solutions to the following (non-linear) differential equation
\be
\label{Vfromf_generic}
V(\phi) \ = \ - f(\phi)^{2} \ + \ f^{\prime}(\phi)^{2} \ .
\ee
Note that since this condition is of the form of a HJ equation \eqref{HJEqforf}, every local solution thereof will define a fake superpotential and
 hence a static HJ flow. 
Through the existence of such an $f$, the extremum $\phi_0$ acquires a fake-superymmetric interpretation since it is characterized by being an extremum
 of $f$, as well as of the potential itself.

Our ODE may be rewritten as 
\be
\label{Vfromf_generic_squareroot}
f^{\prime}(\phi) \ = \ \pm \underbrace{\,\sqrt{V(\phi) \, + \, f^{2}} \,}_{\equiv\,F(\phi,f)} \ .
\ee
Now we can invoke the Cauchy theorem for local existence and uniqueness of solutions, which may be applied any time an initial condition is given
at a point $\left(\phi_{0},f(\phi_{0})\right)$ around which the rhs function $F$ is \emph{Lipschitz}. This implies that there is actually just one
local solution in the neighborhood of a non-critical point, for each choice of sign for $f'$.

However, a particularly interesting situation occurs once an initial condition is given at a critical point, such as $\phi_0$, \emph{i.e.} 
\be
f(\phi_{0}) \ = \ \sqrt{-V(\phi_{0})} \ \equiv \ \sqrt{-V_{0}} \ .
\ee
In the neighborhood of $\phi_0$, $F$ is no longer Lipschitz, thus violating the Cauchy theorem for local existence and uniqueness of solutions to our
ODE. Indeed, there turn out to exist two inequivalent solutions to \eqref{Vfromf_generic_squareroot} which locally start from $\phi_0$. 
The emergent two branches of solutions may be seen by setting up a perturbative expansion around the critical point $\phi_{0}$, 
\be
f_{\textrm{pert.}}(\phi) \ = \ \sum\limits_{k=0}^{\infty}\frac{1}{k!}f^{(k)}(\phi_{0})\,\left(\phi\,-\,\phi_{0}\right)^{k} \ ,
\ee
where the coefficients of the Taylor expansion can be determined by solving the following algebraic system
\be\label{pert_system_f_1D}
\begin{array}{lclc}
V^{(0)}(\phi_{0}) & = & -\,f^{(0)}(\phi_{0})^{2} \, + \, \,f^{(1)}(\phi_{0})^{2} & , \\[2mm]
V^{(1)}(\phi_{0}) & = & -2\,f^{(0)}(\phi_{0})\,f^{(1)}(\phi_{0}) \, + \, 2\,f^{(1)}(\phi_{0})\,f^{(2)}(\phi_{0}) & , \\[2mm]
V^{(2)}(\phi_{0}) & = & -2\,f^{(1)}(\phi_{0})^{2} \,-\,2\,f^{(0)}(\phi_{0})\,f^{(2)}(\phi_{0})\, + \, 
2\,f^{(2)}(\phi_{0})^{2} \, + \, 2\,f^{(1)}(\phi_{0})\,f^{(3)}(\phi_{0})& , \\[2mm]
 & \vdots & & 
\end{array}
\ee
By further setting $f^{(1)}(\phi_{0})\,=\,0$, the above system can be solved order by order separately with, in particular, each derivative of $f$
appearing only linearly except for the second one. The two independent branches are therefore labelled by the two possible choices of $f^{(2)}(\phi_{0})$.

Up to second order in perturbation theory, this yields
\be
\begin{array}{lclclc}
f^{(0)}(\phi_{0}) \, = \, \sqrt{-V_{0}} & , & f^{(1)}(\phi_{0}) \, = \, 0 & , & 
f^{(2)}(\phi_{0}) \, = \, \dfrac{1}{2}\,\left(\sqrt{-V_{0}}\,\pm\,\sqrt{-V_{0}\,+\,2\,V^{(2)}(\phi_{0})}\right) & ,
\end{array}
\ee
where it is worth noticing that the above roots of the second-degree equation for $f^{(2)}(\phi_{0})$ are \emph{only} real if the BF bound is satisfied.
In the next subsection we will combine all of these observations into a set of ``crossing rules'' that will determine possible 
topological obstructions to extending a local solution for $f$ at a global level, thus possibly spoiling the validity of the positive energy theorem.

Before we move to the study of different cases in the AdS landscape, let us generalize the above analysis to include the case of non-static HJ
 flows. As discussed in detail in appendix~\ref{appendix:HJ_flows}, solving the HJ equation for $q_0\,\neq\,0$ is a very complicated problem. However,
for the present scope, we are only interested in the effect of turning on a small $q_0\,>\,0$ to deform a static HJ flow. This, in some sense, 
corresponds to adopting the thin wall approximation, where the thickness of the wall is much smaller compared to the radius of the bubble. This 
assumption seems to be reasonable since the latter is proportional to $q_{0}^{-1/2}$, and will therefore be huge in the $q_0\,\rightarrow\,0$ limit. 

This amounts to taking $a\,=\,0$ inside \eqref{HJ_non_sep}, which yields the following modification of \eqref{Vfromf_generic_squareroot} 
\be
\label{Vfromf_generic_q0}
V(\phi) \ = \ - f(\phi)^{2} \ + \ f^{\prime}(\phi)^{2} \ + \ q_{0} \ ,
\ee
where $q_{0}$ is a positive and small constant. The solution to the $q_{0}$-deformed version of the pertubative system \eqref{pert_system_f_1D} around the critical point $\phi_{0}$ reads
\be
\begin{array}{lc}
f^{(0)}(\phi_{0}) \, = \, \sqrt{-V_{0}\,+\,q_{0}} \ , \qquad  f^{(1)}(\phi_{0}) \, = \, 0 & ,  \\[2mm]
f^{(2)}(\phi_{0}) \, = \, \dfrac{1}{2}\,\left(\sqrt{-V_{0}\,+\,q_{0}}\,\pm\,\sqrt{-V_{0}\,+\,q_{0}\,+\,2\,V^{(2)}(\phi_{0})}\right) & ,
\end{array}
\ee
which, in the $q_0\,\rightarrow\,0$ limit, behaves as
\be
\begin{array}{lc}
f^{(0)}(\phi_{0}) \, = \, \sqrt{-V_{0}}\,+\,\delta f^{(0)} \ , \qquad  f^{(1)}(\phi_{0}) \, = \, 0 & ,  \\[2mm]
f^{(2)}(\phi_{0}) \, = \, \dfrac{1}{2}\,\left(\sqrt{-V_{0}}\,\pm\,\sqrt{-V_{0}\,+\,2\,V^{(2)}(\phi_{0})}\right) \,+\,\delta f^{(2)}  & ,
\end{array}
\ee
where 
\be
\begin{array}{lc}
\delta f^{(0)} \ = \ \dfrac{1}{2 \sqrt{-V_{0}}}\,q_{0}\,+\,\dots & , \\[2mm]
\delta f^{(2)} \ = \ \dfrac{1}{4} \,\left(\dfrac{1}{\sqrt{2 \,V^{(2)}(\phi_{0})\,-\,V_{0}}}\,\pm\,\dfrac{1}{\sqrt{-V_{0}}}\right)\,q_{0}\,+\,\dots & ,
\end{array}
\ee
which manifestly shows that branches with positive and negative second derivative, respectively, tend to be repelled when turning on $q_{0}$.

\subsection*{A ``zoo'' of possible situations in the AdS landscape}

In the previous subsection we have analyzed all the relevant issues that allow us to discuss and classify the different situations in the AdS landscape by means of fake superpotentials and positive
energy theorems. For the present purpose, it may turn out to be useful to spell out the following set of ``crossing rules'' for local solutions to the differential equation \eqref{Vfromf_generic_squareroot}:
\begin{enumerate}
\item At a perturbatively stable critical point of $V$, say $\phi_{0}$, there always exist two branches of local $f$'s. In particular, if we name them $f_{(+)}$ and $f_{(-)}$, 
\begin{itemize}
\item $f_{(\pm)}^{(2)}(\phi_{0})$ are both positive if $\phi_{0}$ is a local maximum,
\item $f_{(\pm)}^{(2)}(\phi_{0})$ have opposite signs if $\phi_{0}$ is a local minimum,
\end{itemize}
\item Two different local branches $f_{1}$ \& $f_{2}$ can only cross at a non-critical point $\phi_{1}$ if 
\be
f_{1}'(\phi_{1}) \ = \ - f_{2}'(\phi_{1}) \ ,
\nonumber
\ee
\item All local solutions are necessarily monotonic and hence they end where $f'$ vanishes. In particular, if this happens away from a local extremum of $V$, say at point $\phi_{1}$, the local branch  
will have a singularity exactly where $V\,=\,-f^{2}$,
\item When turning on $q_{0}\,>\,0$, local branches with positive $f^{(2)}$ will be lifted, while those with negative $f^{(2)}$ will be lowered further.
\end{enumerate} 
An exhaustive classification of all possible situations occuring in the AdS landscape can be obtained by analyzing pairs of critical points of the potential obeying the BF bound. From each critical point
there will be two local branches of fake superpotentials, whose global properties are determined case by case by the above crossing rules. 

\begin{figure}[h!]
\begin{center}
\scalebox{1}[1]{
\begin{tabular}{ccc}
{\bf Situation 1a} & {\bf Situation 1b} & {\bf Situation 1c} \\[3mm]
\includegraphics[scale=0.4,keepaspectratio=true]{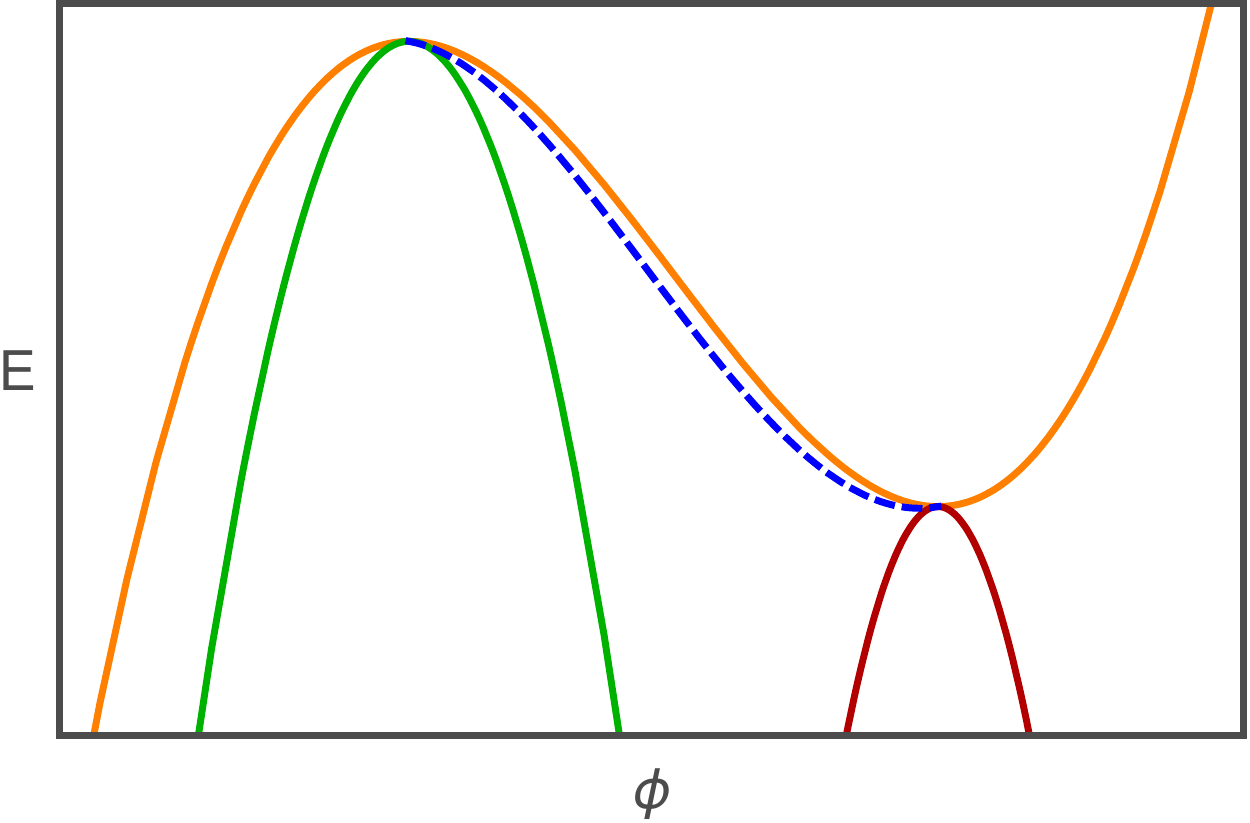} & \includegraphics[scale=0.4,keepaspectratio=true]{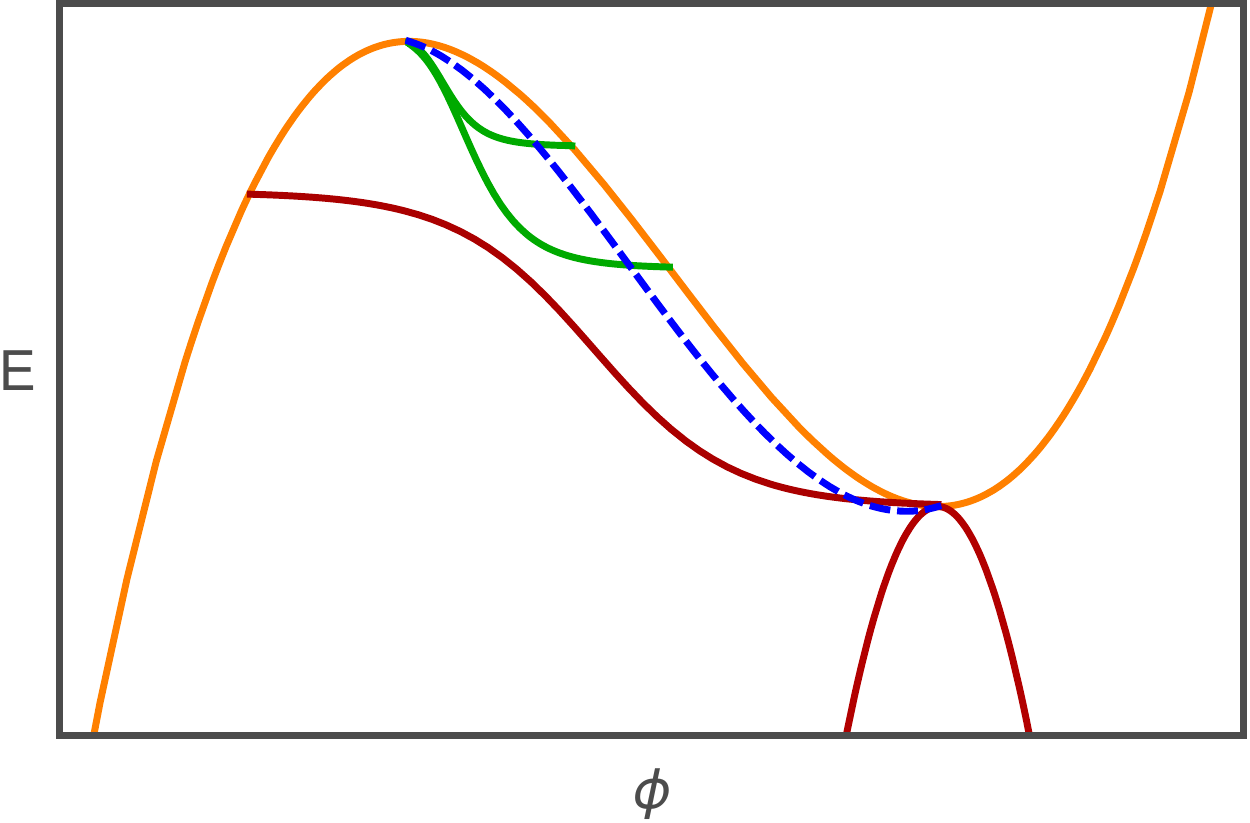} & \includegraphics[scale=0.4,keepaspectratio=true]{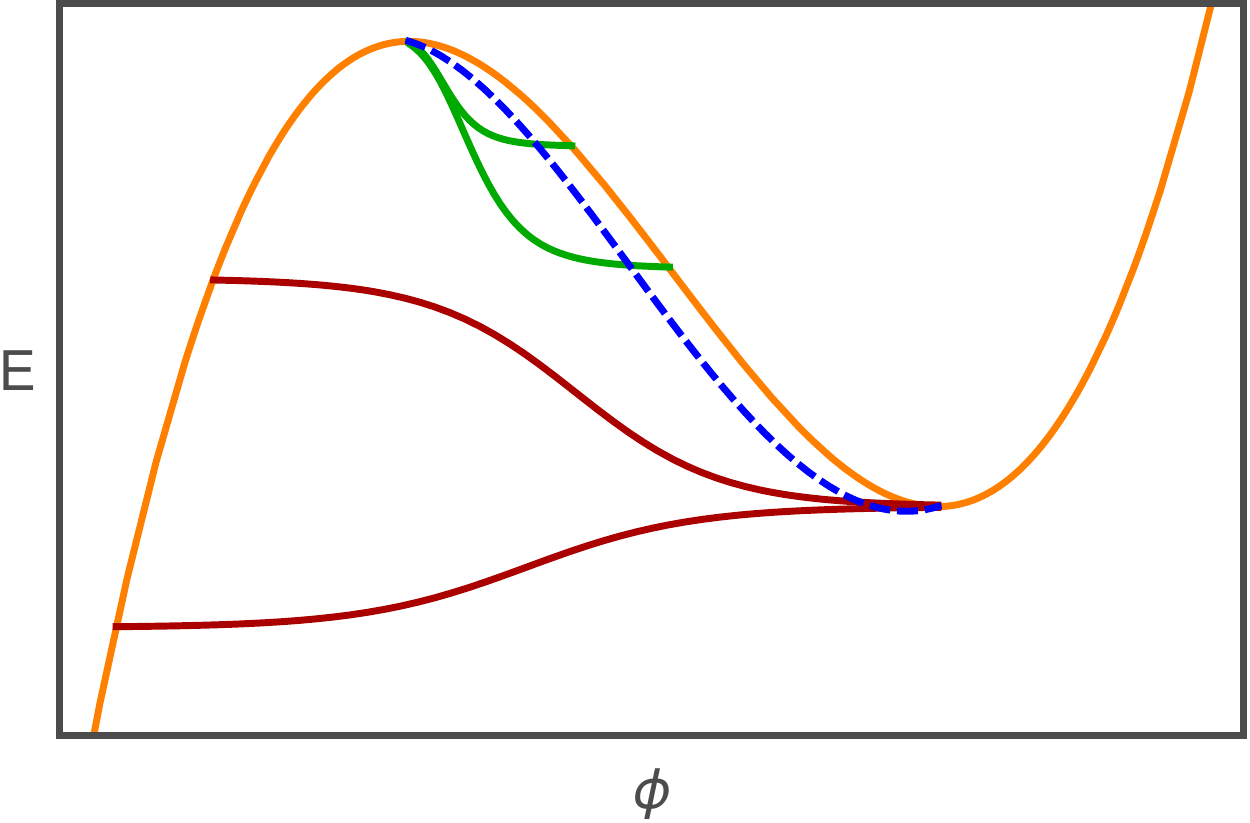}
\end{tabular}
}
\caption{{\it The three different situations that may occur in presence of a local AdS maximum and a local AdS minum.  
\emph{Left:} Both points obey the positive energy theorem and are hence stable against bubble nucleation.  
\emph{Middle:} The maximum turns unstable and decays towards the minimum, which instead stays non-perturbatively stable.
\emph{Right:} Both the maximum and the minimum exhibit non-perturbative instabilities.}}
\label{fig:AdSlandscape_1}
\end{center}
\end{figure}

Let us first consider the case where we have a local maximum and a local minimum of the potential. In such a case, the trajectory of the potential connecting the two extrema happens to be \emph{monotonic}.
The three different possibilities which may occur in this case depend on whether both points, just one or even none admit a globally bounding function obeying the hypothesis of the positive energy 
theorem. These situations are sketched in figure~\ref{fig:AdSlandscape_1}.

\begin{itemize}
\item {\bf Situation 1a:} Both the maximum and the minimum admit globally bounding functions, $-f_{1}^{2}$ \& $-f_{2}^{2}$, respectively drawn in green and brown. They both correspond to the local 
branch labelled by ``$(+)$''. The ``$(-)$'' local branches starting from each of the two points are then topologically forced to meet and merge into a global solution in the interval 
$\left[\phi_{1},\,\phi_{2}\right]$. This solution, which is represented by the blue dotted line, defines a static DW separating the two vacua. Note that such a wall is always extremal and therefore its
tension saturates the bound \eqref{CDL_bound}, thus ruling out tunneling. This conclusion is in perfect agreement with the positive energy theorem.
\item {\bf Situation 1b:} Only the minimum admits a gobally bounding function, represented by one of the two curves in brown and precisely, the one corresponding to the ``$(+)$'' choice. The other local 
branch departing from the minimum has to go up, until it breaks down by hitting the profile of the potential at a non-critical point after going past the maximum. On the other hand, the local branches
starting from the maximum, which are represented by the green curves, both break down by hitting the potential before making it to the minimum. This shows that there are no static DW's in between.
However, if we were now to turn on a $q_{0}\,>\,0$ to search for non-static solutions, the ``$(+)$'' green branch and the ``$(-)$'' brown one would tend to come closer to each other due to rule nr.~4.
As a consequence, there will exist a finite value of $q_{0}$ for which the two vacua are connected. Such non-static solution describes an \emph{ultra-extremal} wall, thus implying graviational tunneling
from the maximum to the minimum through true-vacuum bubble nucleation. 
\item {\bf Situation 1c:} Here neither of the two local extrema possesses a globally bounding function. The discussion concerning the non-perturbative decay of the maximum into the minimum is identical to
situation 1b. The only difference with 1b is that there at least the minimum is stable thanks to the positive energy theorem. Now, since the theorem can no longer be used, there is potentially room for
non-perturbative decay of the minimum towards $-\infty$. Indeed, the expectations are that there should exist a non-static solution flowing from the minimum down to $-\infty$, since the ``$(+)$'' brown
branch is pushed further and further down by turning on $q_{0}\,>\,0$.
\end{itemize}

Let us now move to consider the case of a scalar potential having two local minima. Such a potential profile will then be \emph{non-monotonic} along the path connecting the two extrema.
This implies the existence of a local maximum in between, this giving rise to two different subcases depending on whether or not this maximum satisfies the BF bound. However, if it does, one can just
split the path into two parts in each of which the potential is monotonic\footnote{In such a case, one could have two copies of 1a or 1b. They would both result in a composite wall connecting the two 
minima via the intermediate maximum, thus resulting in a (non-)extremal wall with two insides.} 
and one is back to one of the cases in figure~\ref{fig:AdSlandscape_1}.

\begin{figure}[h!]
\begin{center}
\scalebox{1}[1]{
\begin{tabular}{ccc}
{\bf Situation 2a} & {\bf Situation 2b} & {\bf Situation 2c} \\[3mm]
\includegraphics[scale=0.4,keepaspectratio=true]{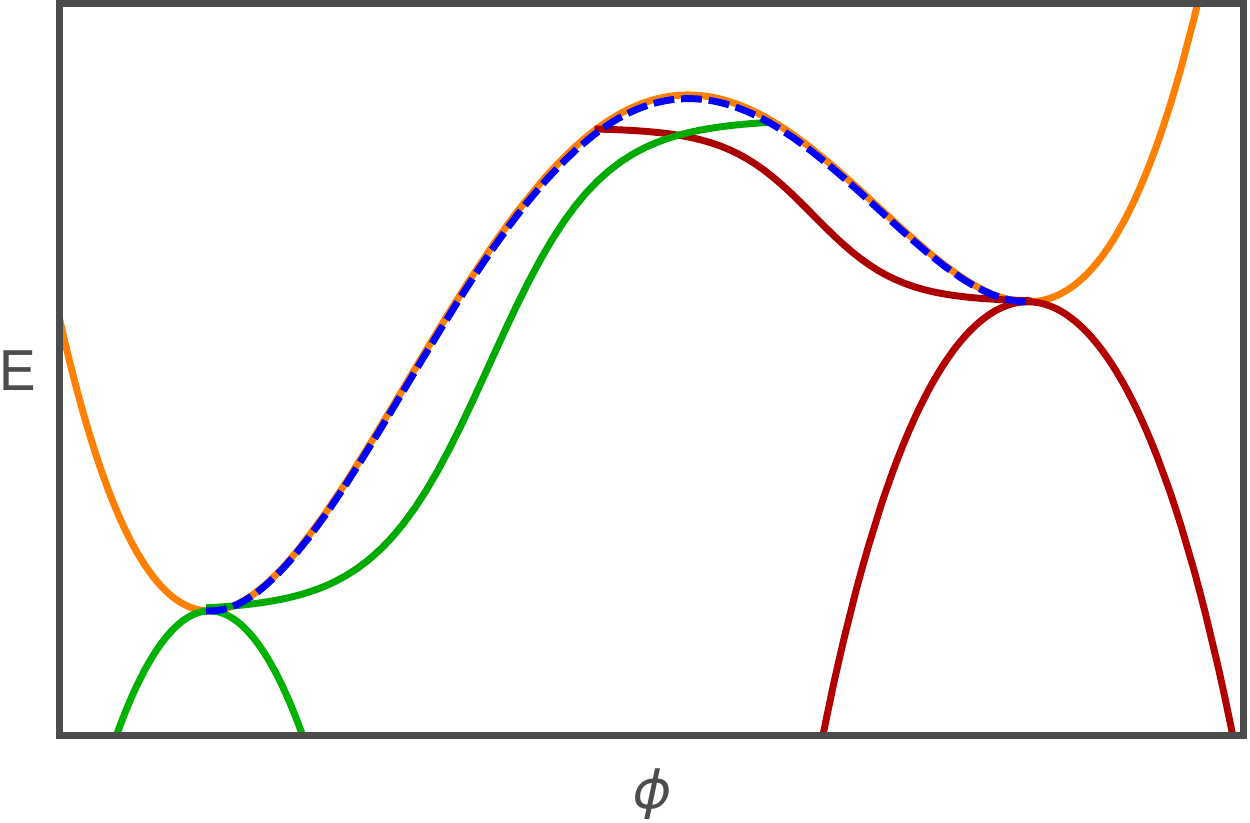} & \includegraphics[scale=0.4,keepaspectratio=true]{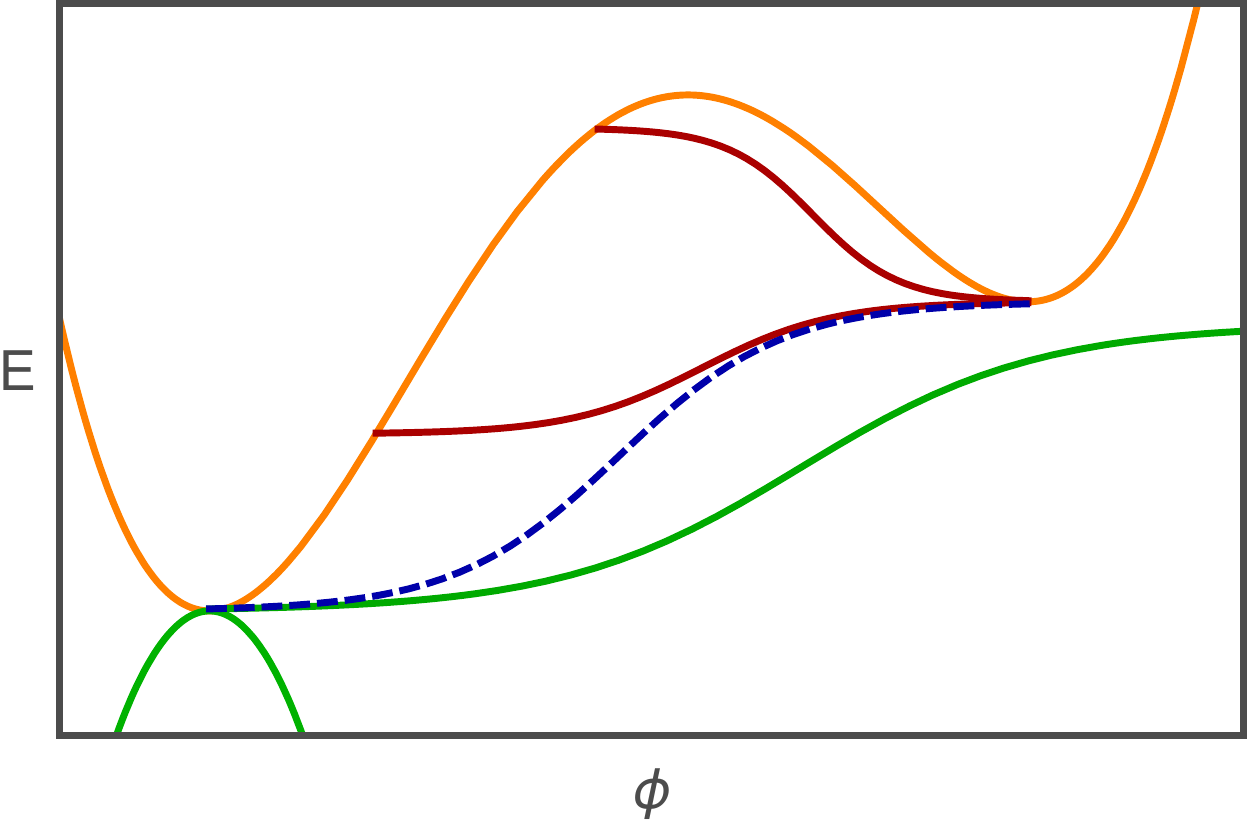} & \includegraphics[scale=0.4,keepaspectratio=true]{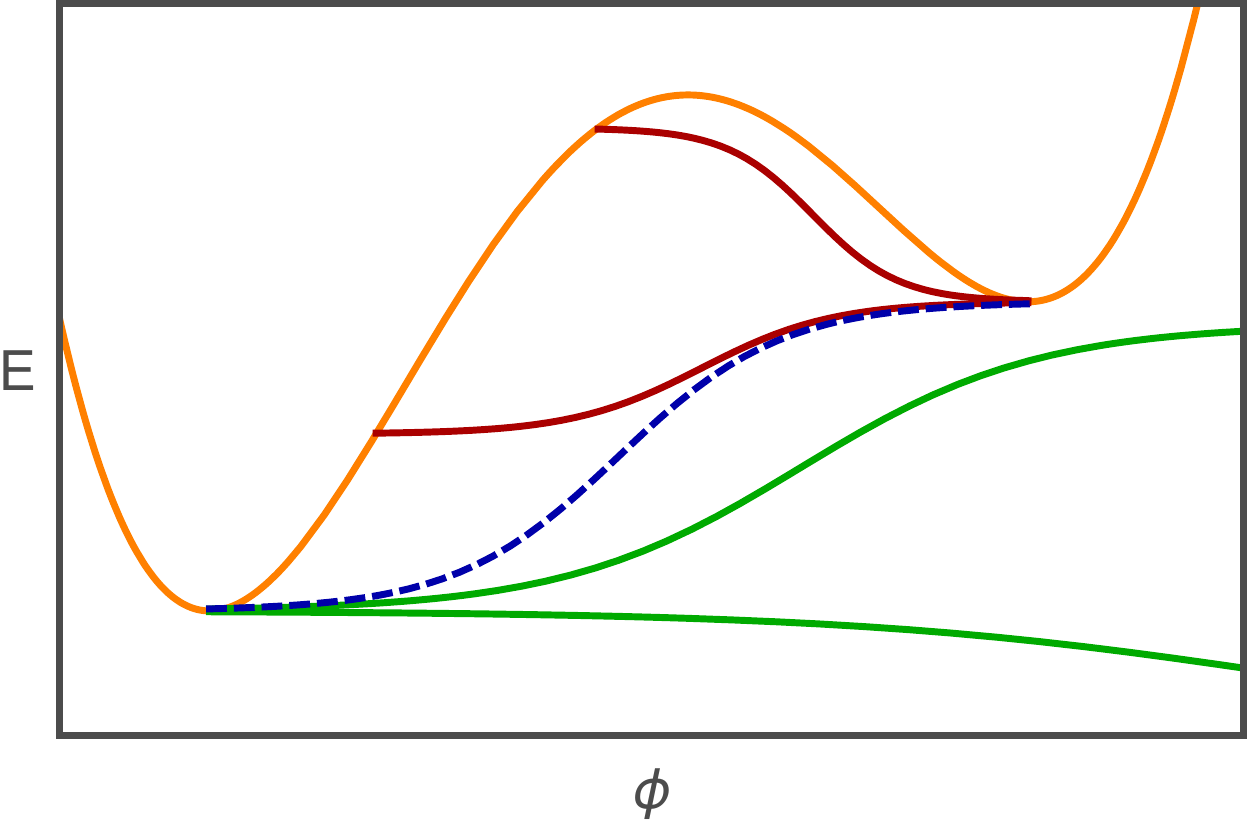}
\end{tabular}
}
\caption{{\it The three different situations that may occur in presence of two local AdS minima separated by an unstable maximum.  
\emph{Left:} Both points obey the positive energy theorem and are hence stable against bubble nucleation.  
\emph{Middle:} The higher minimum turns unstable and decays towards the deeper minimum, which instead stays non-perturbatively stable.
\emph{Right:} Both minima violate the hypothesis of the positive energy theorem and still, the conclusion stays identical to the 2b case.}}
\label{fig:AdSlandscape_2}
\end{center}
\end{figure}

The only truly new case is then the one where the maximum in between violates the BF bound. Also here we have three different situations corresponding to whether both, only one or none of the minima 
admit a globally bounding function. These situations are sketched in figure~\ref{fig:AdSlandscape_2}. 
\begin{itemize}
\item {\bf Situation 2a:} Both minima admit globally bounding functions, $-f_{1}^{2}$ \& $-f_{2}^{2}$, which both correspond to the ``$(+)$'' local branches and are respectively drawn in green and brown.
The ``$(-)$'' local branches starting from each of the two points need to go up but they are not allowed to meet and merge, since this would be against rule nr.~3. They can only cross at one point 
with opposite value of $f'$ as explained in rule nr.~2. In this case, there appears to be no static DW in between and still both vacua are stable against tunneling. In fact, by turning on $q_{0}$, one can 
see that there will be a composite non-static wall passing through the local maximum, where the green and the brown local branches meet with vanishing $f'$. This bubble wall will be \emph{non-extremal},
 its tension strictly respecting the bound \eqref{CDL_bound}, and hence it does not to lead to tunneling.
\item {\bf Situation 2b:} The deeper minimum admits a globally bounding function, whereas the other one does not. In such a situation, the ``$(+)$'' brown local branch hitting the side of the potential
in the static case, approaches the ``$(-)$'' green local branch when $q_{0}\,>\,0$ and they eventually merge to give rise to an \emph{ultra-extremal} bubble wall at finite $q_{0}$. This interpolating
solution describes a gravitational tunneling process from the higher minimum towards the lower and stable one.
\item {\bf Situation 2c:} Here none of the two minima satisfy the hypothesis of the positive energy theorem. This may imply a potential instability of the deeper vacuum as well.
The only practical difference w.r.t. 2b is the possibility for the green branches to hit the potential further on the right, thus creating an available decay channel for the deeper minimum.
\end{itemize}

\section{Two concrete examples}
\label{sec:examples}

In the last section we have discussed the different possibilities that may occur within the AdS landscape on the basis of the set of crossing rules that
we have introduced previously. 
In this section we want to assess to what extent all of the situations encountered there can actually be realized within effective theories that admit
a UV-completion within string theory. To this end, we will present two examples of effective supergravity models coming from string compactifications.
The outcome of our analysis is that none of these fall into the cases where gravitational tunneling occurs.  

\subsection*{Warm-up: massive type IIA on $\textrm{AdS}_{7}\,\times\,S^{3}$}

This class of massive type IIA compactifications is characterized by NS-NS $H_{(3)}$ flux wrapping the internal $S^{3}$ together with spacetime-filling D6-branes localized at the north pole. 
Such string models possess a supersymmetric AdS vacuum, which was first found in \cite{Apruzzi:2013yva} by numerically solving the pure spinor equations and later analytically understood in 
\cite{Apruzzi:2015zna}. These AdS vacua are known to be holographically dual to $(1,0)$ SCFT in six dimensions \cite{Gaiotto:2014lca}. Subsequently in \cite{Passias:2015gya}, these compactifications were shown to admit a consistent truncation yielding minimal 7D gauged supergravity as an effective description.

Such a gauged supergravity description turns out to be a very convenient approach when it comes to searching for solutions.
By adopting this approach, one easily realizes that these theories, besides the aforementioned supersymmetric vacuum, also admit a non-supersymmetric AdS extremum \cite{Dibitetto:2015bia}. 
For the purpose of this section, we will from now on abandon the 10D description in favor of the underlying 7D gauged supergravity formulation. The effective Lagrangian happens to coincide with the one
in \cite{Campos:2000yu} and it reads 
\be
\label{7Daction}
S_{(\textrm{AdS}_{7}\times S^{3})} \ = \ \frac{1}{2\,\kappa_{7}^{2}}\,\int d^{7}x \sqrt{-g} \,\left(\mathcal{R} \, - \, (\partial \phi)^{2} \, - \, 2\,V(\phi)\right) \ ,
\ee
where the scalar potential $V$ is given by 
\be\label{V_ISO3}
V(\phi) \ = \  e^{- \frac{8}{\sqrt{5}}\,\phi} \, \left(4\theta ^2+e^{2\,\sqrt{5}\,\phi} \left(\tilde{q}^2-3 q^2\right)-4 \theta \, e^{\sqrt{5}\,\phi} \,(3q-\tilde{q})\right) \ ,
\ee
where the constants $\theta$, $q$ \& $\tilde{q}$ represent embedding tensor deformation parameters and are related to NS-NS flux, $S^{3}$ extrinsic curvature $\Theta_{ij}$ and Romans' mass, respectively, 
according to the dictionary in table~\ref{table:ET/fluxes7D}.
\begin{table}[h!]
\begin{center}
\begin{tabular}{| c | c | c |}
\hline
IIA fluxes & $\Theta$ components  &  $\mathbb{R}^{+}_{\phi}$ charges \\[1mm]
\hline \hline
$F_{(0)}$ & $\sqrt{2} \, \tilde{q}$ & $+1$ \\[1mm]
\hline
$H_{ijk}$ & $\frac{1}{\sqrt{2}} \, \theta \, \epsilon_{ijk}$ & $-4$ \\[1mm]
\hline
$\Theta_{ij}$ & $q\,\delta_{ij}$ & $+1$ \\[1mm]
\hline
\end{tabular}
\end{center}
\caption{{\it The embedding tensor/fluxes dictionary for the case of massive type IIA reductions on $S^{3}$. The underlying 7D gauging is generically is $\textrm{ISO}(3)$, except when 
$q \, =  \, \tilde{q}$, where it degenerates to $\textrm{SO}(3)$ \protect\cite{Louis:2015mka}.} 
\label{table:ET/fluxes7D}}
\end{table}
By making use of the embedding tensor/fluxes dictionary in \cite{Apruzzi:2016rny}, the scalar potential \eqref{V_ISO3}, which was originally studied in \cite{Campos:2000yu}, is now given a 10D origin.

The scalar potential \eqref{V_ISO3} may written in terms of a \emph{superpotential} as
\be
\label{Vfromf_7D}
V(\phi) \ = \ -\frac{3}{5} \, W(\phi)^{2} \ + \ \frac{1}{2} \, W^{\prime}(\phi)^{2} \ ,
\ee
where
\be
\label{W7D}
W(\phi) \ = \ 2\,\theta\,e^{-\frac{4}{\sqrt{5}}\,\phi} \ + \ e^{\frac{\phi}{\sqrt{5}}}\,\left(3\,q\,-\,\tilde{q}\right) \ .
\ee
The above scalar potential has two AdS critical points when choosing $\textrm{SO}(3)$ as a gauge group, \emph{i.e.}
\be
\begin{array}{lccclc}
\theta \ = \ \dfrac{\lambda}{4} & & , & & q \ = \ \tilde{q} \ = \ \lambda & ,
\end{array}
\ee
one of which is supersymmetric. The relevant physical features of those critical points are summarized in table~\ref{table:AdS7}.
\begin{table}[h!]
\begin{center}
\begin{tabular}{| c || c || c | c | c | c |}
\hline
\textrm{ID} & $\phi_{0}$ & $V_{0}$ & mass spectrum & SUSY & Stability \\[1mm]
\hline \hline
1 & $0$ & $-\frac{15}{4}\,\lambda^{2}$ &
$\begin{array}{cc}0 & (\times\,3)\\[1mm] \boldsymbol{-\frac{8}{15}} & (\times\,1)\\[1mm] 
\frac{16}{15} & (\times\,5)\\[1mm] \frac{8}{3} & (\times\,1) \end{array}$ & \checkmark & \checkmark \\[1mm]
\hline
2 & $-\frac{\log 2}{\sqrt{5}}$ & $-\frac{5}{2^{2/5}}\,\lambda^{2}$ &
$\begin{array}{cc}0 & (\times\,8)\\[1mm] \boldsymbol{\frac{4}{5}} & (\times\,1)\\[1mm] 
\frac{12}{5} & (\times\,1)\end{array}$ & $\times$ & \checkmark \\[1mm]
\hline
\end{tabular}
\end{center}
\caption{{\it The two AdS solutions of minimal gauged supergravity in $D=7$ admitting massive type IIA on AdS$_{7}\times S^{3}$ as 10D interpretation. 
The mass spectra include nine extra scalar modes sitting in the three vector multiplets that contain all closed string excitations, while $m^{2}_{\phi}$ is marked in bold.} 
\label{table:AdS7}}
\end{table}

As generically already argued in the previous section, in order to discuss the possibility of gravitational tunneling between Sol.~1 \& and 2 in table~\ref{table:AdS7}, we need to search for interpolating 
\emph{static} flow solutions. To this end, we make use of the usual flat wall \emph{Ansatz} for the 7D metric and for the scalar $\phi$
\be
\label{DW7DAnsatz}
\left\{
\begin{array}{lclc}
ds_{7}^{2} & = & dz^{2} \ + \ e^{2\,a(z)} \, ds_{\textrm{Mkw}_{6}}^{2} & , \\[2mm]
\phi & = & \phi(z) & .
\end{array}
\right.
\ee
By plugging \eqref{DW7DAnsatz} into the action \eqref{7Daction}, one finds
\be
\label{1Daction}
S_{(1\textrm{D})} \ \overset{\textrm{up to bdy}}{=} \ \frac{1}{\kappa_{7}^{2}}\,\int dz \, e^{6a} \left(15\,\left(a^{\prime}\right)^{2} \, - \, \frac{1}{2}\,\left(\phi^{\prime}\right)^{2}\, - \, V(\phi)\right) \ ,
\ee
where $^{\prime}$ denotes a derivative w.r.t. the $z$ coordinate. The above action implies the following second-order field equations
\be
\label{2ndorder_7D}
\left\{
\begin{array}{lclc}
15 \, \left(a^{\prime}\right)^{2} \, - \, \dfrac{1}{2} \, \left(\phi^{\prime}\right)^{2} \, + \, V & = & 0 & , \\[2mm]
\phi^{\prime\prime} \, + \, 6 \, a^{\prime}\,\phi^{\prime} \, - \, \partial_{\phi}V & = & 0 & ,
\end{array}
\right.
\ee
The corresponding interpolating solution between Sol.~1 \& 2 can be found by making use of the Hamilton-Jacobi (HJ) formalism (see 
appendix~\ref{appendix:HJ_flows}). This procedure yields the following first-order reformulation of \eqref{2ndorder_7D} \cite{Campos:2000yu}
\be
\label{1storder_7D}
\left\{
\begin{array}{lclc}
a^{\prime} & = & \dfrac{1}{5} \, f & , \\[2mm]
\phi^{\prime} & = & -\, \partial_{\phi}f & ,
\end{array}
\right.
\ee
provided that the functional $f(\phi)$ satisfy the following non-linear ODE
\be
\label{Vfromf_7D}
V(\phi) \ = \ -\frac{3}{5} \, f(\phi)^{2} \ + \ \frac{1}{2} \, f^{\prime}(\phi)^{2} \ .
\ee
Note that the above ODE has an obvious global solution given by the supersymmetric superpotential in \eqref{W7D}, \emph{i.e.} $f_{\textrm{SUSY}}(\phi) \, = \, W(\phi)$.
In this case, the equations \eqref{1storder_7D} define a BPS flow that can only hit supersymmetric critical points.
Any other local solution of \eqref{Vfromf_7D} would define a fake superpotential and hence an extremal but non-BPS flow that can connect any pair of perturbatively stable AdS critical points. 
By means of this $f$, these critical points gain a fake-supersymmetric interpretation.

As already explained in the previous section, by assigning an initial condition at one of the two critical points, \emph{i.e.}
\be
f(\phi_{0}) \ = \ \sqrt{-\frac{5}{3}\,V_{0}} \ ,
\ee
our ODE \eqref{Vfromf_7D} fails to obey the local uniqueness theorem and it admits two branches of local solutions (see crossing rule nr.~1).
Indeed, by solving the perturbative system \eqref{pert_system_f_1D} at second order for both critical points in table~\ref{table:AdS7}, one finds
\be\label{Nilsson_branch_SUSY}
\begin{array}{lclclc}
f^{(0)}(0) \, = \, \frac{5}{2} & , & f^{(1)}(0) \, = \, 0 & , & f^{(2)}_{(\mp)}(0) \, = \, \left\{\begin{array}{l}1 \\[1mm]
\frac{1}{2} \end{array}\right. & ,
\end{array}
\ee
for the supersymmetric solution labelled by ``1'', and
\be
\begin{array}{lclclc}
f^{(0)}(-\frac{\log 2}{\sqrt{5}}) \, = \, \frac{5}{2^{1/5}\sqrt{3}} & , & f^{(1)}(-\frac{\log 2}{\sqrt{5}}) \, = \, 0 & , & f^{(2)}_{(\mp)}(-\frac{\log 2}{\sqrt{5}}) \, = \, 
\left\{\begin{array}{l}-\frac{\sqrt{7}\,-\,\sqrt{3}}{2^{1/5}} \\[1mm]
\frac{\sqrt{7}\,+\,\sqrt{3}}{2^{1/5}} \end{array}\right. & ,
\end{array}
\ee
for the non supersymmetric extremum labelled by ``2''. In both cases, branch ``$-$'' turns out be a global solution which defines a notion of
positive energy by providing a global bound of the form
\be
V(\phi) \ \geq \ -\frac{3}{5}\,f(\phi)^{2} \ , \qquad \forall \,\, \phi\, \in \, \mathbb{R} \ ,
\ee
thus verifying the hypothesis of the positive energy theorem.
In particular, the one constructed perturbatively from the supersymmetric vacuum exactly coincides with the superpotential in \eqref{W7D}. The other branches turn out to connect smoothly and define a global function on the interval $\left(-\frac{\log 2}{\sqrt{5}},\,0\right)$. Such a function exactly coincides with the fake superpotential defining the static DW connecting Sol.~1 \& 2. The remarkable feature of this non-BPS static DW is that its associated fake superpotential happens to be non-analytical at $\phi\,=\,0$, where $f^{(3)}$ becomes infinite. However, such a divergence is still such
that $f^{(1)}(0)\,f^{(3)}(0)\,\rightarrow\,0$.

\begin{figure}[h!]
\begin{center}
\scalebox{1}[1]{
\begin{tabular}{ccc}
\includegraphics[scale=1,keepaspectratio=true]{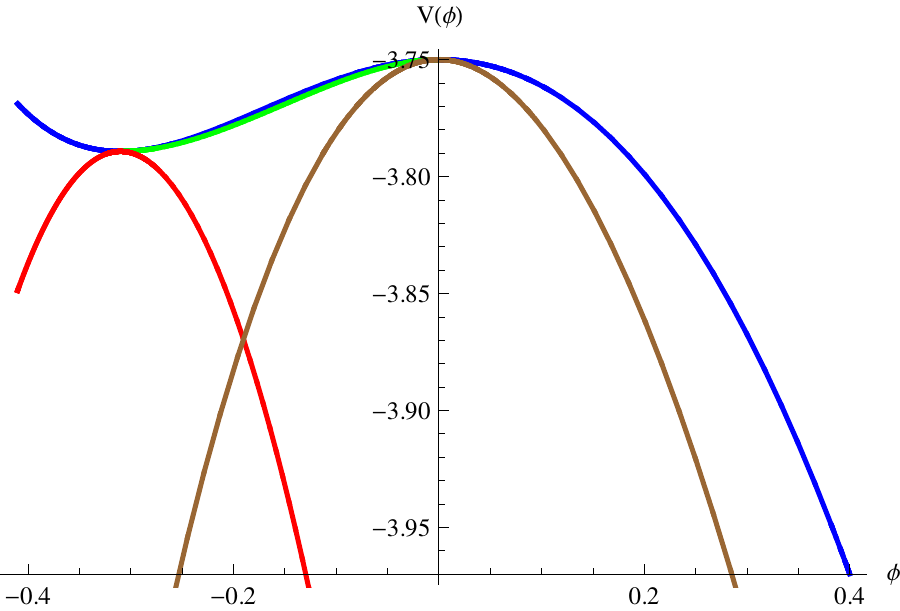} &  & \includegraphics[scale=0.6,keepaspectratio=true]{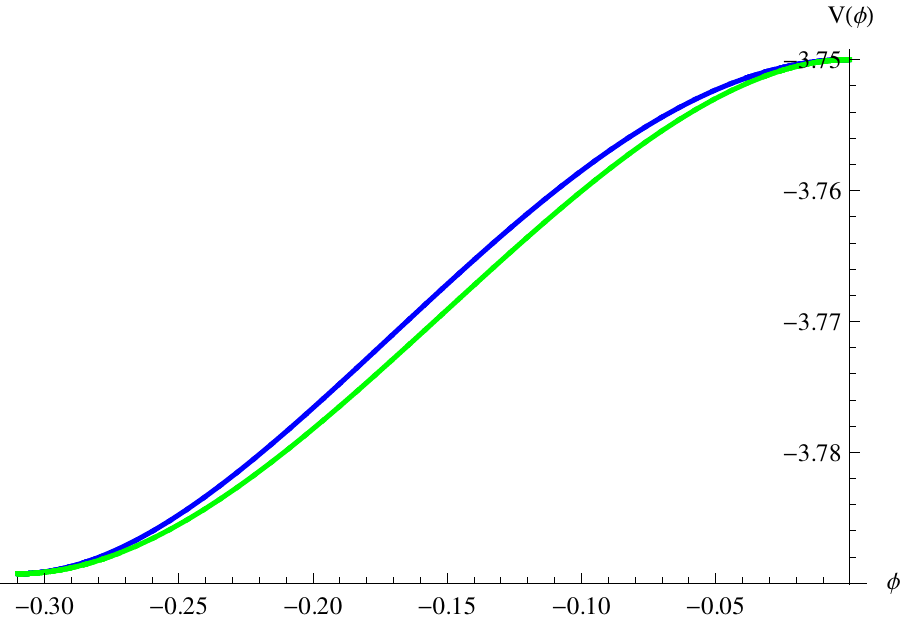} \\[-25mm]
 & $\rightarrow$ & \\[20mm]
\end{tabular}
}
\caption{{\it The non-perturbative stability of massive type IIA on $\textrm{AdS}_{7}\times S^{3}$ models summarized. The blue curve shows the profile
of the scalar potential \eqref{V_ISO3}, with a non-supersymmetric local minimum (left) and a supersymmetric local maximum (right).
From both points there starts a globally bounding function $-\frac{3}{5}\,f^2$ ensuring their non-perturbative stability (curves in red \& brown,
 respectively). Note that these branches only cross at one point and respecting crossing rule nr.~2. 
This exactly realizes \textbf{Situation~1a} in figure~\ref{fig:AdSlandscape_1}.
Finally, the green curve represents the bounding
function defining the static DW (zoomed in on the right).}}
\label{fig:Nilsson}
\end{center}
\end{figure}

The existence of this static interpolating flow is a direct consequence of the presence of the two global bounding branches. 
Indeed these other branches cannot intersect any of the two global solutions in any point in the interval 
$\left(-\frac{\log 2}{\sqrt{5}},\,0\right)$, due to crossing rules nr.~2 \& 3.
This situation is sketched in figure~\ref{fig:Nilsson}.

One last comment which is worth making concerns the asymptotic behavior of the two globally bounding functions starting from Sol.~1 \& 2 
(the branches drawn in brown and red, respectively). As already mentioned earlier, the globally bounding function around the supersymmetric point is
precisely the superpotential \eqref{W7D} of the theory and it behaves as $e^{\frac{\phi}{\sqrt{5}}}$ as $\phi\,\rightarrow\,+\infty$, \emph{i.e.}
exactly like the square root of the leading term inside the scalar potential at infinity. 

The other global fake superpotential constructed around the non-supersymmetric point exhibits yet a steeper behavior at infinity, \emph{i.e.}
$e^{\sqrt{\frac{6}{5}}\,\phi}$, which is making sure that a further intersection of the curve $-\frac{3}{5}\,f^{2}$ with the potential profile be
 avoided in a neighborhood of $+\infty$. Such an asymptotic behavior arises from a deeply different way of solving the HJ constraint \eqref{Vfromf_7D}
 w.r.t. the supersymmetric one. One may indeed see that $\sqrt{\frac{6}{5}}$ happens to be exactly the critical value of $\mathbb{R}^{+}_{\phi}$ weight
 that arranges for the cancellation between the leading terms coming from $f^{2}$ \& $f'^{2}$, such in a way that the corresponding term is absent in 
$V$. 

Note that the above peculiarity will actually not occur at $-\infty$, where both global $f$'s behave as $e^{-\frac{4}{\sqrt{5}}\,\phi}$, since this
time $-\frac{4}{\sqrt{5}} \, < \, -\sqrt{\frac{6}{5}}$, thus making the latter contribution subleading. However, the asymptotic analysis on the this
other side is somewhat less relevant since what prevents the crossing anyway from happening is the potential going positive and even asymptotically
approaching $+\infty$.

The above analysis shows that the two curves, which are respectively drawn in brown and red in figure~\ref{fig:Nilsson}, both satisfy the hypothesis
of the positive energy theorem. Moreover, as a consequence of our crossing rules, one gets for free the existence of the interpolating static DW which
contains very valuable information from a holographic viewpoint \cite{Apruzzi:2016rny}. 
This completes the proof of the impossibility of gravitational tunneling through spontaneous bubble nucleation within the AdS landscape of
$S^3$ compactifications of massive type IIA string theory.

\subsection*{A multi-field case: massive type IIA on $\textrm{AdS}_{4}\,\times\,S^{3}\,\times\,S^{3}$}

We would like now to consider a more involved situation featuring more scalar fields. To this end, we move to a class of compactifications of massive
 type IIA supergravity down to 4D. The explicit case of interest to us is that of $\textrm{AdS}_{4}\,\times\,S^{3}\,\times\,S^{3}$ massive IIA backgrounds. These compactifications are supported by NS-NS $H_{(3)}$ flux as well as R-R $F_{(0)}$, $F_{(2)}$, $F_{(4)}$ and $F_{(6)}$ fluxes wrapping 
cycles in the internal manifold $S^{3}\,\times\,S^{3}$, which is thought of as a group manifold whose curvature is parametrized by metric flux $\omega$.
No extended D-brane or O-plane sources turn out to be needed in order to solve the 10D field equations.

These compactifications are known to admit a supersymmetric AdS vacuum, which was found by using many different approaches, ranging from 
 $\textrm{SU}(3)$-structures to STU-models (see \emph{e.g.} refs~\cite{Derendinger:2004jn,Villadoro:2005cu,Aldazabal:2007sn,Caviezel:2008ik}). Due to the complete absence of local sources, these models also admit an 
$\mathcal{N}=8$ gauged supergravity description including all $70$ scalar fields of the maximal theory. The corresponding embedding tensor/fluxes
 dictionary was derived in \cite{Dibitetto:2012ia}, where all the quadratic constraints required for the consistency of the gauging of the maximal theory where interpreted as the absence of different types of BPS branes.

However, if one restricts to the $\textrm{SO}(3)$-invariant sector of the maximal supergravity theory, these compactifications can be effectively
 described by an $\mathcal{N}=1$ supergravity in $D=4$ coupled to three chiral multiplets, each of which contains one complex scalar field (usually named $(S,T,U)\,\equiv\,\Phi^{i}$) 
spanning an $\textrm{SL}(2,\mathbb{R})/\textrm{SO}(2)$ coset. Such a minimal supergravity model is also known as an STU-model. 
Upon performing the above truncation, the corresponding invariant embedding tensor components give rise to flux-induced superpotential couplings.
The obtained superpotential reads
\be
W(\Phi) \ = \ a_{0} \, - \, 3\,a_{1}\,U \, + \, 3\,a_{2}\,U^{2} \, - \, a_{3}\,U^{3} \, - \, b_{0}\,S \, + \, 3\,b_{1}\,S\,U \, 
+ \, 3\, c_{0}\,T \, + \,3\, \left(2c_{1}\,-\,\tilde{c}_{1}\right)\,T\,U  ,
\ee
where the above couplings are related to type IIA fluxes through the dictionary reported in table~\ref{table:ET/fluxes4D}.
\begin{table}[h!]
\begin{center}
\begin{tabular}{| c | c | c |}
\hline
IIA fluxes & $W$ couplings  &  STU charges \\[1mm]
\hline \hline
$F_{(0)}$ & $a_{3}$ & $(+\frac{1}{2};\,+\frac{3}{2};\,-\frac{3}{2})$ \\[1mm]
\hline
$F_{ai}$ & $a_{2}$ & $(+\frac{1}{2};\,+\frac{3}{2};\,-\frac{1}{2})$ \\[1mm]
\hline
$F_{aibj}$ & $a_{1}$ & $(+\frac{1}{2};\,+\frac{3}{2};\,+\frac{1}{2})$ \\[1mm]
\hline
$F_{aibjck}$ & $a_{0}$ & $(+\frac{1}{2};\,+\frac{3}{2};\,+\frac{3}{2})$ \\[1mm]
\hline
$H_{ijk}$ & $b_{0}$ & $(-\frac{1}{2};\,+\frac{3}{2};\,+\frac{3}{2})$ \\[1mm]
\hline
$H_{abk}$ & $c_{0}$ & $(+\frac{1}{2};\,+\frac{1}{2};\,+\frac{3}{2})$ \\[1mm]
\hline
${\omega_{ij}}^{c}$ & $b_{1}$ & $(-\frac{1}{2};\,+\frac{3}{2};\,+\frac{1}{2})$ \\[1mm]
\hline
${\omega_{ab}}^{c}$ & $\tilde{c}_{1}$ & $(+\frac{1}{2};\,+\frac{1}{2};\,+\frac{1}{2})$ \\[1mm]
\hline
${\omega_{aj}}^{k}$ & $c_{1}$ & $(+\frac{1}{2};\,+\frac{1}{2};\,+\frac{1}{2})$ \\[1mm]
\hline
\end{tabular}
\end{center}
\caption{{\it The embedding tensor/fluxes dictionary for the case of massive type IIA reductions on $S^{3}\times S^{3}$. The labels ``$abc$'' \& ``$ijk$'' respectively refer to internal directions on the 
two different $S^{3}$ factors. The above superpotential deformations were originally identified in \protect\cite{Dall'Agata:2009gv} with those type IIA fluxes which are even w.r.t. a 
$\mathbb{Z}_{2}$-involution defined by spacetime-filling O6-planes which further wrap $S^{3}_{a}$. Later, in \protect\cite{Dibitetto:2014sfa}, this dictionary was completed by also including 
the orientifold-odd sector.} 
\label{table:ET/fluxes4D}}
\end{table}

The 4D effective Lagrangian reads
\be
\label{4Daction}
S_{(\textrm{AdS}_{4}\times S^{3}\times S^{3})} \ = \ \frac{1}{2\,\kappa_{4}^{2}}\,\int d^{4}x \sqrt{-g} \,\left(\mathcal{R} \, - \, 
K_{i\bar{j}}\,(\partial \Phi^{i})(\partial \Phi^{\bar{j}}) \, - \, 2\,V(\Phi,\bar{\Phi})\right) \ ,
\ee
where the kinetic metric is defined by $K_{i\bar{j}}\,\equiv\,\partial_{i}\partial_{\bar{j}}K$, and the K\"ahler potential reads
\be
K(\Phi,\bar{\Phi}) \ = \ -\log\left(-i\,(S-\bar{S})\right)\, -\,3\,\log\left(-i\,(T-\bar{T})\right)\,-\,3\,\log\left(-i\,(U-\bar{U})\right) \ .
\ee
Finally, the scalar potential is determined as
\be
\label{VfromW_4D}
V(\Phi,\bar{\Phi}) \ = \ e^{K(\Phi,\bar{\Phi})}\,\left(-3\,|W(\Phi)|^{2}\,+\,|D_{\Phi}W|^{2}\right) \ ,
\ee
where $D$ denotes the K\"ahler-covariant derivative.

The set of extrema of the effective scalar potential was exhaustively studied in \cite{Dibitetto:2011gm}. The resulting landscape consisted of purely
AdS vacua, which could be viewed as four different critical points of the same potential corresponding to the following flux choice
\be
\begin{array}{cccccccc}
a_{0} \ = \ \frac{3}{2}\,\sqrt{10}\,\lambda & , & a_{1} \ = \ \frac{1}{2}\,\sqrt{6}\,\lambda & , & a_{2} \ = \ -\frac{1}{6}\,\sqrt{10}\,\lambda & , & a_{3} \ = \ \frac{5}{6}\,\sqrt{6}\,\lambda & , \\[2mm]
b_{0} \ = \ -\frac{1}{3}\,\sqrt{6}\,\lambda & , & b_{1} \ = \ \frac{1}{3}\,\sqrt{10}\,\lambda & , & c_{0} \ = \ \frac{1}{3}\,\sqrt{6}\,\lambda & , & c_{1} \ = \ \tilde{c}_{1} \ = \ \sqrt{10}\,\lambda & , 
\end{array}
\ee
one of which is the supersymmetric vacuum. The relevant physical features of these critical points are summarized in table~\ref{table:AdS4}.
\begin{table}[h!]
\begin{center}
\begin{tabular}{| c || c || c | c | c | c |}
\hline
\textrm{ID} & $\left(S_{0},\,T_{0},\,U_{0}\right)$ & $V_{0}$ & mass spectrum & SUSY & Stability \\[1mm]
\hline \hline
1 & $\left(\begin{array}{c}i\\i\\i\end{array}\right)$ & $-\lambda^{2}$ &
$\begin{array}{cc}0 & (\times\,1)\\[1mm]-\frac{2}{3} & (\times\,1)\\[1mm] \frac{1}{3}\left(4\pm\sqrt{6}\right) & (\times\,1)\\[1mm]\frac{1}{9}\left(47\pm\sqrt{159}\right) & (\times\,1)\end{array}$
 & \checkmark & \checkmark \\[1mm]
\hline
2 & $\left(\begin{array}{c}\frac{4}{\sqrt{3}}\left(\frac{1}{\sqrt{5}}+\frac{2^{1/3}}{5^{2/3}}\,i\right)\\ \frac{4\,\cdot\,2^{1/3}}{\sqrt{3}\,5^{2/3}}\,i\\ 
\frac{1}{\sqrt{3}}\left(-\frac{1}{\sqrt{5}}+\frac{2\,\cdot\,2^{2/3}}{5^{1/3}}\,i\right)\end{array}\right)$ & 
$-\frac{125 \sqrt{3} \,5^{2/3}}{512 \,\cdot\,2^{1/3}}\,\lambda^{2}$ &
$\begin{array}{cc}0 & (\times\,1)\\[1mm] -\frac{4}{5} & (\times\,1)\\[1mm] -\frac{2}{5} & (\times\,1)\\[1mm] 2 & (\times\,1)\\[1mm] \frac{64}{15} & (\times\,1)\\[1mm] \frac{20}{3} & (\times\,1)
\end{array}$ & $\times$ & $\times$ \\[1mm]
\hline
3 & $\left(\begin{array}{c}\frac{2}{\sqrt{5}}\left(2\sqrt{3}+\,i\right)\\ \frac{2}{\sqrt{5}}\,i\\ \frac{1}{\sqrt{5}}\left(-\sqrt{3}+2\,i\right)\end{array}\right)$ & $-\frac{25}{48} \sqrt{5} \,\lambda^{2}$ &
$\begin{array}{cc}0 & (\times\,2)\\[1mm] 2 & (\times\,2)\\[1mm] \frac{20}{3} & (\times\,2)\end{array}$ & $\times$ & \checkmark \\[1mm]
\hline
4 & $\left(\begin{array}{c}\frac{4}{\sqrt{3}}\left(\frac{1}{\sqrt{5}}+\frac{2^{1/3}}{5^{1/6}}\,i\right)\\ \frac{4\,\cdot\,2^{1/3}}{3\,\sqrt{3}\,5^{1/6}}\,i\\ 
\frac{1}{\sqrt{3}}\left(-\frac{1}{\sqrt{5}}+\frac{2\,\cdot\,2^{2/3}}{5^{1/3}}\,i\right)\end{array}\right)$ & $-\frac{135 \sqrt{3} \,5^{2/3}}{512 \,\cdot\,2^{1/3}}\,\lambda^{2}$ &
$\begin{array}{cc}0 & (\times\,2)\\[1mm] \frac{4}{3} & (\times\,1)\\[1mm] 2 & (\times\,1)\\[1mm] 6 & (\times\,1)\\[1mm] \frac{20}{3} & (\times\,1)\end{array}$ & $\times$ & \checkmark \\[1mm]
\hline
\end{tabular}
\end{center}
\caption{{\it The four AdS solutions of minimal STU models in $D=4$ admitting massive type IIA on AdS$_{4}\times S^{3}\times S^{3}$ as 10D interpretation. 
The mass spectra including all 70 scalar modes sitting in the $\mathcal{N}=8$ gravity multiplet that contain all closed string excitations where found
in \protect\cite{Dibitetto:2012ia}, while here we only report the masses of the 6 real modes within the STU sector.} 
\label{table:AdS4}}
\end{table}

While Sol.~2 is perturbatively unstable due to the presence of a mode below the BF bound, the other two non-supersymmetric extrema appear perturbatively
stable within the STU sector. Therefore it makes sense to address the issue of their non-perturbative stability against quantum tunneling. 
As already widely discussed, this is intimately connected with the existence of fake superpotentials and static HJ flows.
In order to make contact with the formalism introduced in appendix~\ref{appendix:HJ_flows} for a multi-field case, we first reformulate the STU-model
described by the action \eqref{4Daction} in terms of six real scalars $\phi^{I}\,\equiv\,\{\sigma,\,s,\,A,\,B,\mu,\,u\}$, where
\be
\begin{array}{lclclc}
S \ = \ s \, + \, i \, \sigma & , & T \ = \ B \, + \, i \, A & , & U \ = \ u \, + \, i \, \mu & .
\end{array}
\ee
In terms of the above real fields, the condition \eqref{VfromW_4D} becomes 
\be
\label{VfromW_4D}
V(\phi) \ = \ -3\,{\cal W}(\phi)^{2}\,+\,2\,K^{I J} \, \frac{\partial {\cal W}}{\partial \phi^I} \, \frac{\partial {\cal W}}{\partial \phi^J} \ ,
\ee
where ${\cal W}\ \equiv \ e^{K/2}\,|W|$ \cite{Cvetic:1992bf}.
Now we are ready to start looking for interpolating static solutions. To this end, we cast the \emph{Ansatz}
\be
\label{DW4DAnsatz}
\left\{
\begin{array}{lclc}
ds_{4}^{2} & = & dz^{2} \ + \ e^{2\,a(z)} \, ds_{\textrm{Mkw}_{3}}^{2} & , \\[2mm]
\phi^{I} & = & \phi^{I}(z) & .
\end{array}
\right.
\ee
By plugging \eqref{DW4DAnsatz} into the action \eqref{4Daction} and applying the HJ formalism as explained in appendix~\ref{appendix:HJ_flows}, one
 finds that the second-order equations of motion for the six scalars and the function $a(z)$ are equivalent to the first-order flow equations in
\eqref{HJ_q0=0}, supplemented with the HJ constraint \eqref{HJEqforf}. Note that an obvious global solution to the PDE \eqref{HJEqforf} is given by 
the real superpotential of the model, \emph{i.e.} $f_{\textrm{SUSY}}\,=\,{\cal W}$, whereas any other local solution thereof can be interpreted as a
 fake superpotential.

In order for gravitational tunneling to be excluded for the non-supersymmetric and nevertheless stable solutions in table~\ref{table:AdS4}, we would
 need to find the existence of globally bounding functions verifying the positive energy theorem. Once such objects are found, we will investigate
whether or not this is paired up with the existence of static interpolating DW's. This will help us identify which of the situations described in
 section~\ref{sec:DW} are actually realized within this model.  

\begin{figure}[h]
\begin{center}
\includegraphics[scale=0.45]{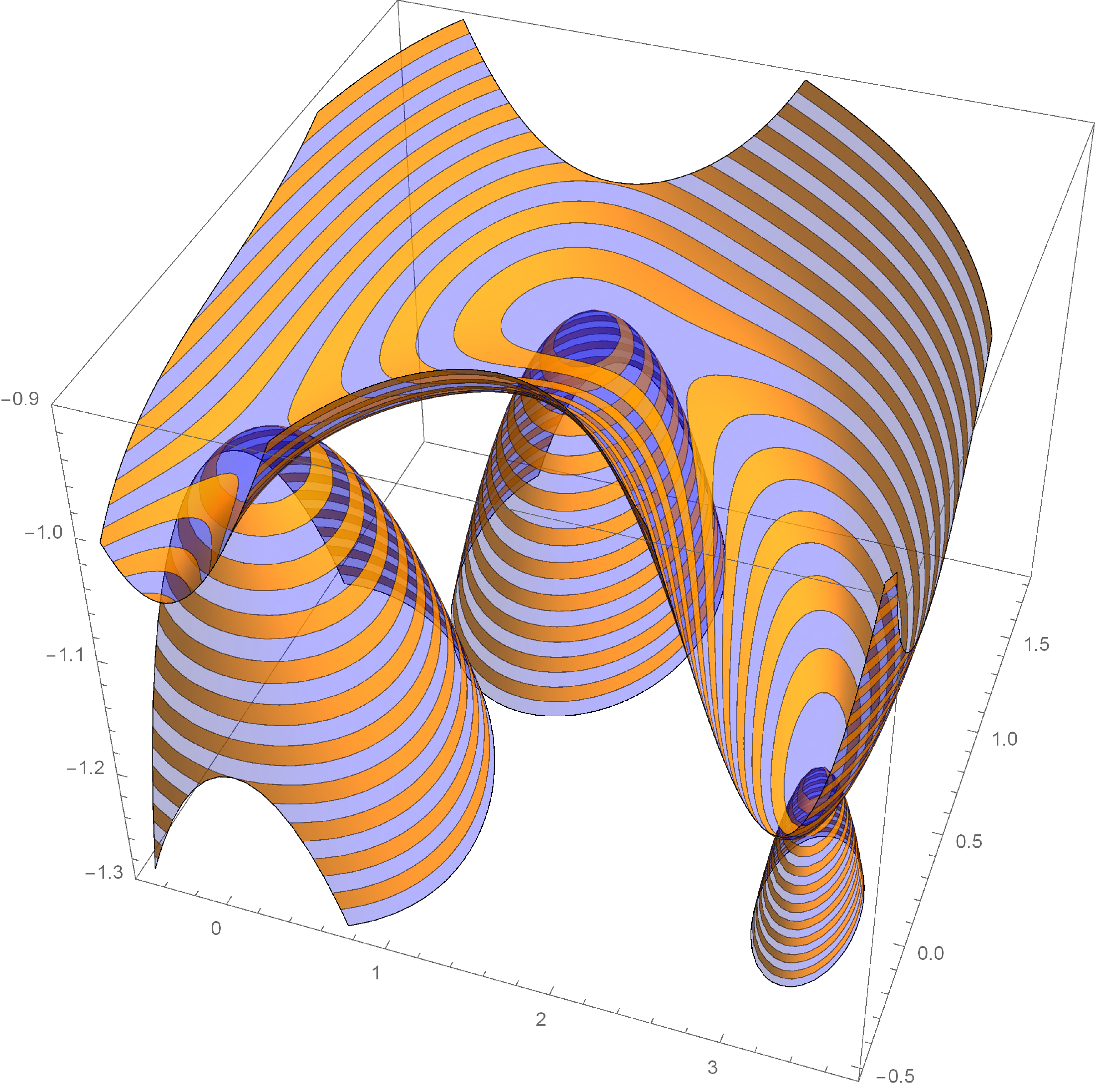}
\caption{\it The non-perturbative stability of massive type IIA on $\textrm{AdS}_{4}\times S^{3}\times S^{3}$ models summarized. The above sheet represents the profile of the scalar potential \eqref{V_ISO3} in a particular two-dimensional slice of the scalar manifold, with a supersymmetric local
 extremum on the left and two additional non-supersymmetric ones (Sol.~3 \& 4 in table~\ref{table:AdS4}).
From all points there starts a globally bounding function $-3\,f^2$ ensuring their non-perturbative stability (represented by the paraboloids peaked
at each critical point).}
\label{fig:Stability_6Scalars}
\end{center}
\end{figure}

We were actually able to solve the PDE \eqref{HJEqforf} to determine the globally bounding functions around all three critical point in table~\ref{table:AdS4} respecting the BF bound, \emph{i.e.} Sol.~1,
3 \& 4. The numerical technique is based on the generalization of the perturbative method sketched in section~\ref{sec:DW} to the case of six scalars. Some technical aspects concerning the method are
collected and summarized in appendix~\ref{appendix:solving_PDEs}. In order to obtain numerically satisfactory solutions in this context we have used perturbation theory up to order $15$.
The potential profile together with the three different globally bounding functions are plotted on a two-dimensional sheet in figure~\ref{fig:Stability_6Scalars}.
It is worth mentioning that the aforementioned functions verifying the positive energy theorem for all three critical points correspond to the following choice of local branches\footnote{As explained in 
detail in appendix~\ref{appendix:solving_PDEs}, there exist $2^{6}\,=\,64$ inequivalent local branches at every critical point, labelled by the 
eigenvalues of $f^{(2)}(\phi_{0})$. For the sake of simplicity, here we just give their signs.}
\be
\begin{array}{lcclc}
f^{(2)}(\phi_{1}) \ \sim \ \textrm{diag}(+ \, + \, + \, - \, -\, 0) & , & \textrm{and } & f^{(2)}(\phi_{3,4}) \ \sim \ \textrm{diag}(+ \, + \, + \, + \, +\, +) & ,
\end{array}
\ee
where the choice for Sol.~1 exactly coincides with $f_{\textrm{SUSY}}$. 

\begin{figure}[h]
\begin{center}
\includegraphics[scale=1]{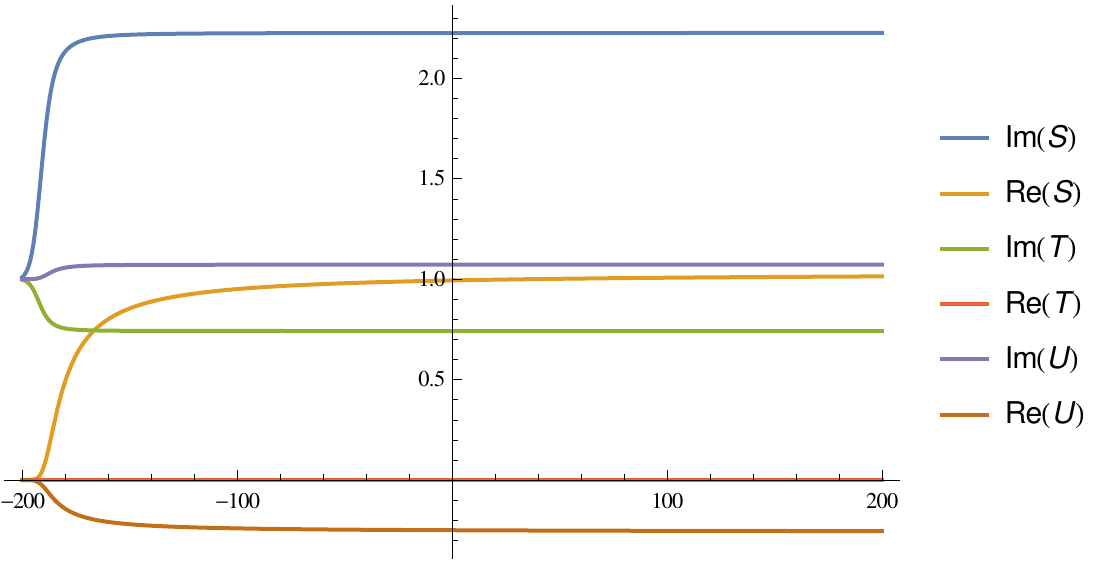}
\caption{\it The profile of the six real scalars in our STU-model along the flow representing the static DW interpolating between Sol.~4 \& Sol.~1.}
\label{fig:DW_flow}
\end{center}
\end{figure}

Moreover, there turns out to exist a path connecting Sol.~1 \& 4 along which the potential is monotonic. 
In fact, an order $15$ perturbative expansion around $\phi_{4}$ with a different choice of local branch with $f^{(2)}(\phi_{4}) \ \sim \ \textrm{diag}(- \, - \, - \, - \, 0\, 0)$ turns out to define a 
static DW solution, which is necessarily extremal but non-BPS. Such an object solves the HJ flow equations in \eqref{HJ_q0=0_new} determined by the above local branch for $f$ as an HJ generating 
functional. The profile of the six real scalars $\{\sigma,\,s,\,A,\,B,\mu,\,u\}$ in the model is plotted along the flow in figure~\ref{fig:DW_flow}. 

\begin{figure}[h!]
\begin{center}
\scalebox{1}[1]{
\begin{tabular}{cc}
\includegraphics[scale=0.47,keepaspectratio=true]{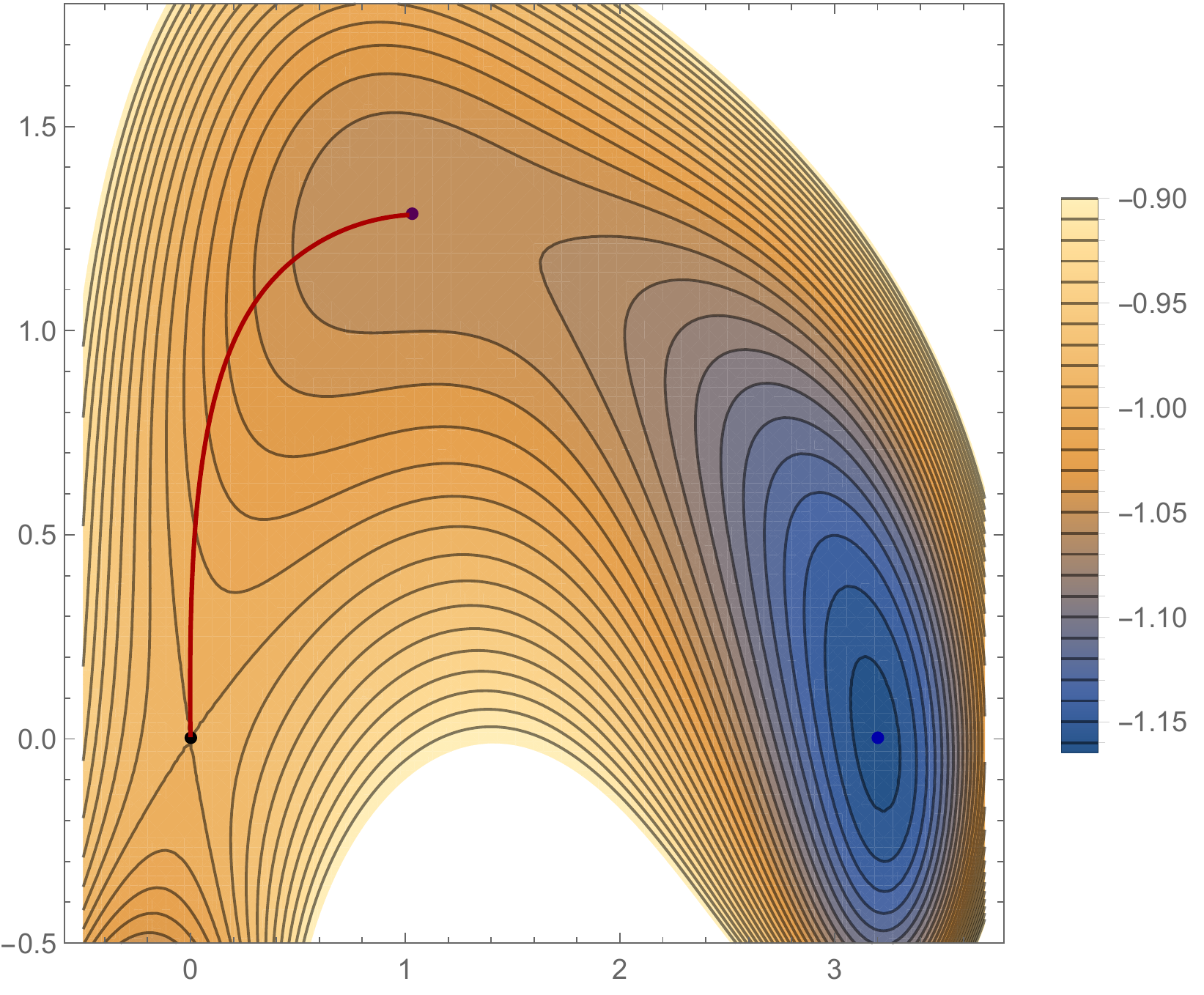} &  \includegraphics[scale=0.47,keepaspectratio=true]{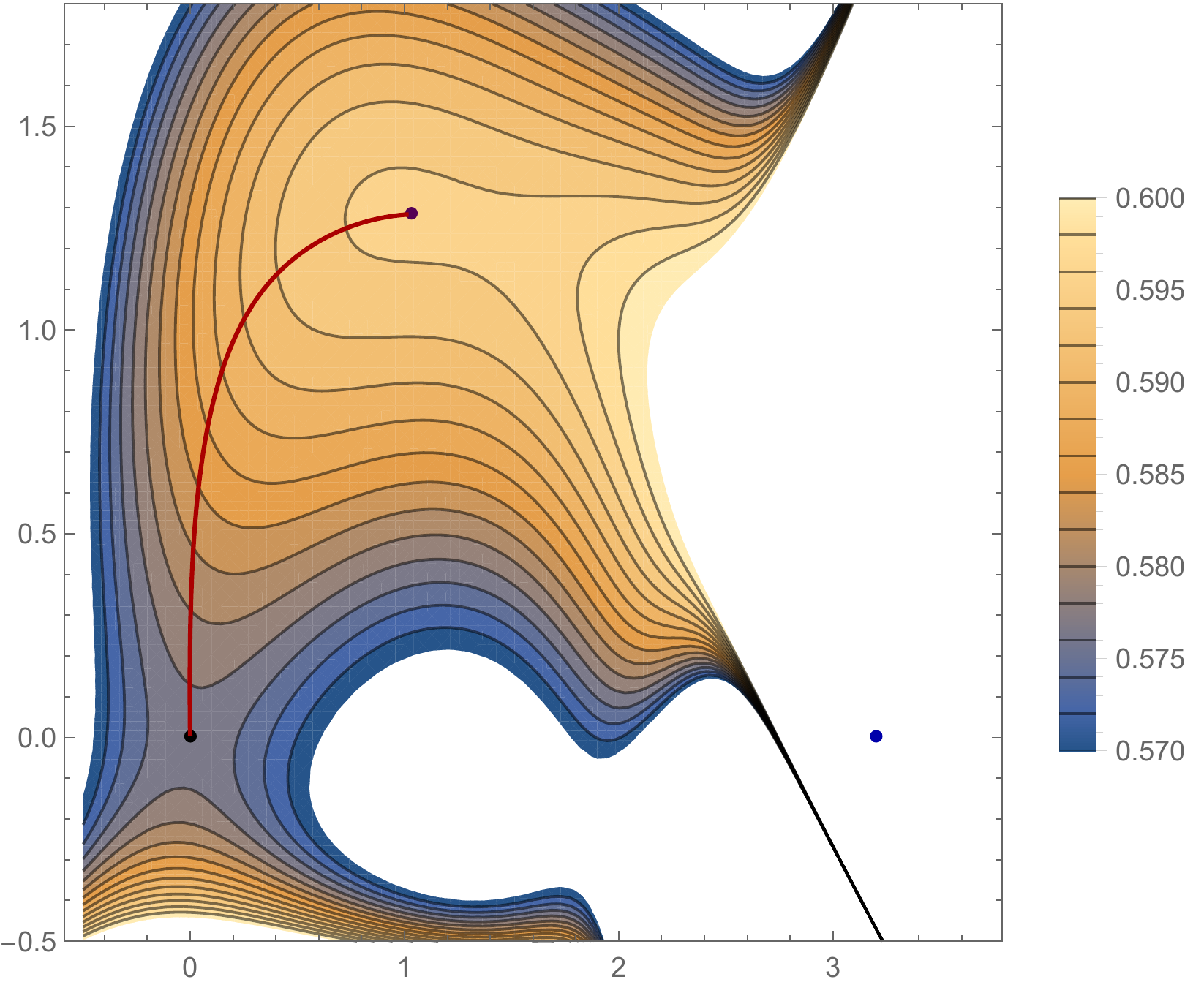} 
\end{tabular}
}
\caption{{\it The above diagrams represent a plane within the scalar manifold containing Sol.~1 (lower-left), 3 (lower-right) \& 4 (upper-middle) of table~\ref{table:AdS4}.  
\emph{Left:} The level curves of the scalar potential \eqref{VfromW_4D}.  
\emph{Right:} The level curves of the fake superpotential $f$ defining the static DW.
In both plots the thick red line denotes the projection of the DW curve connecting Sol.~4 to 1. Note that, since such curve is just a projection of the actual extremal trajectory, it needs not be 
orthogonal to the level curves of $f$.}}
\label{fig:Level_curves}
\end{center}
\end{figure}

Figure~\ref{fig:Level_curves} instead shows the level curves of $V(\phi)$ \& $f(\phi)$, respectively, as well as the projection of the DW curve within the plane containing Sol.~1, 3 \& 4. 
A particularly stricking feature of the level curves of the fake superpotential plotted on the rhs of figure~\ref{fig:Level_curves} is that they blow up exactly on the right boundary of the colored 
region. That curve is expected to be the envelope of all points where the flows indentified by that local branch of $f$ obtained by varying the intial condition hit the side of the potential and cause a 
breakdown of the perturbative expansion. Interestingly, this happens before reaching critical point nr.~3.

\begin{figure}[h!]
\begin{center}
\scalebox{1}[1]{
\includegraphics[scale=0.8,keepaspectratio=true]{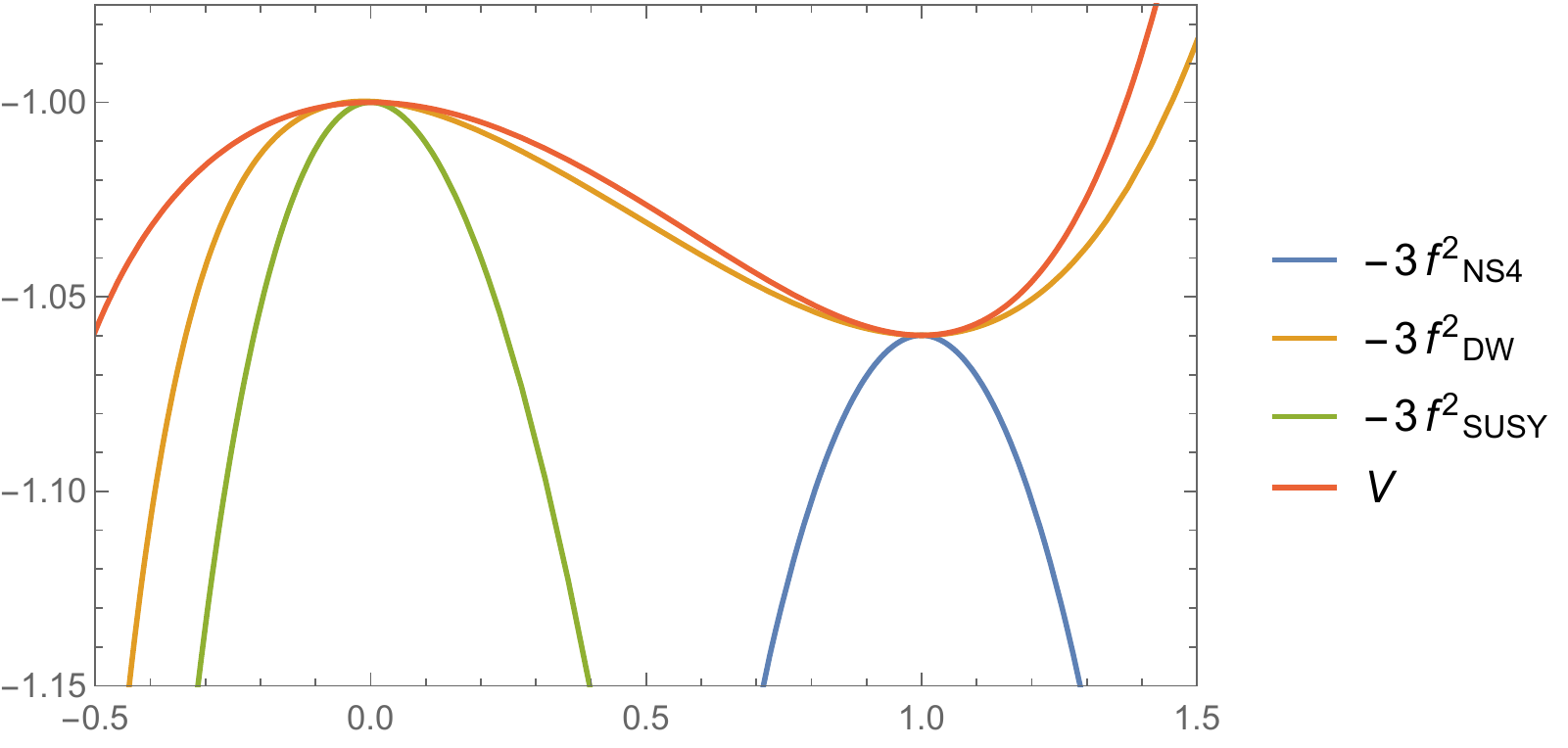}
}
\caption{{\it The potential profile (red), the globally bounding functions $-3\,f_{\textrm{SUSY}}^{2}$ (green) \& $-3\,f_{\xcancel{\textrm{SUSY}}}^{2}$ (blue), and finally $-3\,f_{\textrm{DW}}^{2}$ 
(orange). This situation exactly reproduces \textbf{Situation~1a} in  figure~\ref{fig:AdSlandscape_1}.
The asymptotic behavior of $f_{\textrm{SUSY}}^{2}$ vs $f^{2}_{\xcancel{\textrm{SUSY}}}$ as $\sigma\,=\,\textrm{Re}(S)\,\rightarrow\,0^{+}$; the former diverges exactly as $V$ itself, whereas the latter has a faster
slope, \emph{i.e.} $-\sqrt{3}$, but exactly such that it cancels against $(\partial f)^{2}$ when calculating $V$. Similar arguments correctly predict the asymptotic behavior at other boundaries of 
$\mathcal{M}_{\textrm{scalar}}$.}}
\label{fig:1Dsection_1}
\end{center}
\end{figure}

The above analysis provides the needed evidence that quantum graviational tunneling is actually forbidden also within the landscape of massive type IIA on $S^{3}\,\times\,S^{3}$. To conclude, we would
like to identify which of the 1D ``cartoon'' situations shown in figure~\ref{fig:AdSlandscape_1} \& \ref{fig:AdSlandscape_2} are concretely realized within this model.
To this end, we must discuss the potential profile together with the relevant local branches of fake superpotentials in a suitable one-dimensional slice of $\mathcal{M}_{\textrm{scalar}}$.
For the pair of critical points 1 \& 4, the situation can be easily visualized along the straight line connecting the two points and the resulting sketch is shown in figure~\ref{fig:1Dsection_1}.
For the other pairs (\emph{i.e.} 1 \& 3, as well as 3 \& 4), the situation becomes a bit more difficult to depict. However, in both cases there is a bump in the middle of the path that prevents a static 
DW from existing and all the local branches going up happen to hit the side of the potential somewhere. The only technical complication that makes it generically hard to visualize is to find the right
1D slice where a given branch hits the potential. However, after finding it, the picture for these two cases would look exactly like \textbf{Situation~2a} in figure~\ref{fig:AdSlandscape_2}.

\section{Conclusions}
\label{sec:conclusions}

In this paper we have studied the conditions that allow for quantum gravitational tunneling between AdS vacua within the context of effective gravity 
coupled to scalar fields. Such a phenomenon can happen through spontaneous nucleation of a true vacuum bubble which starts expanding. 
We have started reviewing some well-known results derived within the thin-wall approximation, \emph{i.e.} neglecting the dynamical evolution of scalars.
This approach allows one to derive a bound on the tension of the wall separating the two vacua.

Subsequently, we have adopted a different approach to the problem which is based on the formulation of a positive energy theorem which turns out to establish the equivalence between non-perturbative stability and fake supersymmetry. This statement has been obtained by means of the Hamilton-Jacobi
formalism that provides a first-order formulation of the studied second-order problem.
The possible existence of fake superpotentials has furthermore allowed us to discuss all possible situations in the AdS landscape, some of which lead to
tunneling. The general expectations turn out to be in line with those of \cite{Harlow:2010az}, where related issues were previously discussed.
It may be worth mentioning that our results, in comparison with \cite{Freedman:2003ax}, in some cases do exclude tunneling even in cases where interpolating static solutions are absent.

Finally, in order to investigate whether the aforementioned situations allowing for tunneling actually occur in a consistent theory of quantum gravity, we have analyzed two explicit examples of effective supergravity theories with multiple AdS vacua which have a stringy origin. In particular, they
describe massive type IIA on $\textrm{AdS}_{7}\,\times\,S^{3}$ and $\textrm{AdS}_{4}\,\times\,S^{3}\,\times\,S^{3}$, respectively.
The result of our analysis in these specific cases is that quantum tunneling is forbidden within these models. In particular, for the latter case, our proof is consistent with results in \cite{Narayan:2010em},
where some explicit non-perturbative decay channels for this model where ruled out.

The same machinery could be in principle used to study similar cases of string compactifications where the AdS landscape is known, like \emph{e.g.}
M-theory on $\textrm{AdS}_{7}\,\times\,S^{4}$ and $\textrm{AdS}_{4}\,\times\,S^{7}$, or the recently discovered trunctation of massive type IIA on 
$\textrm{AdS}_{4}\,\times\,S^{6}$ \cite{Guarino:2015vca}, or type IIB on $\textrm{AdS}_{5}\,\times\,S^{5}$. 
In this last case in fact, the analysis carried out in \cite{Girardello:1998pd} can already be seen as a proof
of absence of quantum tunneling in the light of our results. Many other examples may be added to the above list in the next future.

All of this leads us to conjecture that non-perturbative stability within the AdS landscape might be a universal feature of any consistent
quantum gravity model, rather than being merely a special coincidence associated with the specific properties of the cases analyzed in this work.
We should mention that there are, though, examples in the literature where some models of stringy AdS vacua were found to be non-perturbatively unstable \cite{Horowitz:2007pr,Gaiotto:2009yz}. 
We hope to come back to those examples in the future in order to understand them in the light of our conjecture. Our expectations are that they could possess instabilities within other sectors of the closed
string excitations already at a pertubative level.

As far as holographic applications are concerned, we found that non-perturbative stability of AdS vacua can be either associated with the existence of a
static though non-BPS domain wall (see figure~\ref{fig:Nilsson} \& \ref{fig:1Dsection_1}), or even of a non-extremal one (as expected between Sol.~1 \& 3 or 3 \& 4 of table~\ref{table:AdS4}).
While the holographic interpretation of the former is well-understood in terms of RG flows between (non-)supersymmetric conformal fixed points in the
 dual field theory, such an interpretation remains obscure in the case of the latter. We speculate that these might be related with RG flows at finite
temperature in their holographic dual, but it would be interesting to study this issue in detail in the future.

%
%

\section*{Acknowledgments}

We would like to thank Thomas Van Riet for very stimulating discussions. The work of the authors is supported by the Swedish Research Council (VR).

\newpage

%
%

\appendix

\section{Hamilton-Jacobi flows and AdS (non-)static DW's}\label{appendix:HJ_flows}

In this appendix we review the dynamical equations describing time-(in)dependent flows interpolating between AdS vacua within a theory of gravity coupled to a set of scalar fields 
$\left\{\phi^{I}\right\}_{I\,=\,1,\,\dots\,N}$. We will first cast a time-dependent \emph{Ansatz} describing an expanding bubble and then write down the set of second-order equations describing the
evolution of such a bubble. Subsequently, we will discuss the first-order formulation of the above problem by applying the Hamilton-Jacobi (HJ) formalism. We will specify here to four spacetime dimensions,
but it may be worth mentioning that a similar analysis can be carried out in other $D$. 

\subsection*{The second-order formulation}
The Lagrangian which dynamically describes the coupling of Einstein gravity to the aforementioned scalar fields is given by
\be
\label{Full_4D_action}
S[g_{\mu\nu},\phi^{I}] \ = \ \frac{1}{2\kappa_{4}^{2}}\,\int d^{4}x\,\sqrt{-g}\,\left(\mathcal{R}\,-\,
g^{\mu\nu}\,K_{IJ}\,\partial_{\mu}\phi^{I}\,\partial_{\nu}\phi^{J}\,-\,V(\phi)\right) \ ,
\ee
where $\kappa_{4}$ is related to the 4D Newton's constant and $K_{IJ}$ represents the (non-canonical) kinetic metric.
The metric of an expanding curved bubble reads
\be
\label{bubble_metric_4D}
ds_{4}^{2} \ = \ e^{2a(\zeta)}\, \bigg[d\zeta^{2}\ \underbrace{\, - \, dt^{2} \,+\,S(t)^{2}\,\left(\frac{dr^{2}}{1\,-\,\kappa\,r^{2}} \, + \, r^{2} \,d\varphi^{2}\right)}_{\textrm{FRW}_{3}}\bigg] \ ,
\ee
where $\kappa\,=\,0,\,\pm 1$ is the bubble curvature parameter and $a(\zeta)$ \& $S(t)$ are two unknown functions. Moreover, the scalars are assumed to
 just depend on $\zeta$.

The components of the Einstein tensor evaluted on the background \eqref{bubble_metric_4D} read
\begin{eqnarray}
&& G_{tt} \ = \ \frac{\kappa + \dot{S}^2}{S^2} \ - \ a'^2 \ - \ 2 \ a'' \ ,
\\
&& G_{\zeta\zeta} \ = \ - \frac{\kappa + \dot{S}^2}{S^2} \ - \ 2 \ \frac{\ddot{S}}{S} \ + \ 3 \  a'^2 \ ,
\\
&& G_{rr} \ = \ \frac{S^2}{1 - \kappa r^2} \ \left( a'^2 \ + \ 2 \ a'' \ - \ \frac{\ddot{S}}{S}  \right) \ ,
\\
&& G_{\varphi \varphi} \ = \ S^2 \ r^2 \ \left( a'^2 \ + \ 2 \ a'' \ - \ \frac{\ddot{S}}{S}  \right) \ ,
\end{eqnarray}
where ``$\,\prime$\,'' denotes the derivative w.r.t. the $\zeta$ coordinate, whereas ``$\,\cdot\,$'' denotes the time derivative.
The components of the stress-energy tensor can be calculated through $T_{\mu\nu}\,\equiv\,-\frac{2}{\sqrt{-g}}\,\frac{\delta S}{\delta g^{\mu\nu}}$,
yielding
\begin{eqnarray}
&& T_{tt} \ = \ \frac{1}{2} \ K_{I J} \ \phi'^{I} \ \phi'^{J} \ + \ e^{2a} \ V \ ,
\\
&& T_{\zeta\zeta} \ =  \ \frac{1}{2} \ K_{I J} \ \phi'^{I} \ \phi'^{J} \ - \ e^{2a} \ V \ ,
\\
&& T_{rr} \ = \ - \ \frac{S^2}{1 - \kappa r^2} \ T_{tt} \ ,
\\
&& T_{\varphi \varphi} \ = \ - \ S^2 \ r^2 \ T_{tt} \ .
\end{eqnarray}
The consistency of the full set of Einstein equations requires $G_{rr} \, \overset{!}{=} \, - \, \frac{S^2}{1 - \kappa r^2} \, G_{tt}$, which implies
\begin{equation}
 \frac{\ddot{S}}{S} \ \overset{!}{=} \ \frac{\kappa + \dot{S}^2}{S^2}\ \overset{!}{=} \ \textrm{const.} \ \equiv \ q_0 \ . 
\end{equation}
The static case of a flat DW may be recovered when taking $q_0\,=\,0$, whereas the $q_0\,>\,0$ case turns out to be relevant for discussing expanding
spherical non-extremal bubbles that can lead to gravitational tunneling. 

After imposing the above consistency constraint, the original set of Einstein equations together with the equations of motion for the scalar fields just 
reduces to the following second-order differential problem\footnote{From now on we set $\kappa_4\,=\, 1$.}
\be
\label{DW_2nd_order}
\left\{
\begin{array}{lclc}
3 \,  a'^2 \, - \, 3 \, q_0 \,-\,\frac{1}{2} \, K_{I J} \, \phi'^{I} \, \phi'^{J} \, + \, e^{2a} \, V & = & 0 & , \\[2mm] 
\phi''^{I} \, + \, 2 \, a' \, \phi'^{I} \, + \, {\Gamma^{I}}_{JL} \, \phi'^{J} \, \phi'^{L} \, - \, e^{2a} \, K^{IJ} \, \partial_J \, V & = & 0 & ,
\end{array}\right.
\ee
where ${\Gamma^{I}}_{JL}$ denote the components of the Christoffel connection on the scalar manifold
\begin{equation}
{\Gamma^{I}}_{JL} \ \equiv \ \frac{1}{2} \, K^{IM} \left( \partial_J  K_{LM} \, + \, \partial_L K_{JM} \, - \, \partial_M K_{JL} \right) \ .
\end{equation}

\subsection*{The first-order HJ formulation}

The 4D action \eqref{Full_4D_action} evaluted on the curved-bubble background introduced in \eqref{bubble_metric_4D} reads
\begin{equation}
\label{action_1D_appendix}
S_{(1\textrm{D})} \ = \ \int d\zeta\,e^{2a}\,\left(3\,\left(a'^2\,+\,q_0 \right)\,-\,\frac{1}{2}\,K_{I J}\,\phi'^{I}\,\phi'^{J}\,-\,V\,e^{2a}\right)   
\ .  
\end{equation}
The conjugate momenta to $a$ and $\phi^{I}$ are then given by
\be
\begin{array}{lclclc}
\pi^{(a)} & = & \dfrac{\partial \mathcal{L}}{\partial a'} & = & 6 \, a' \, e^{2a} & , \\[2mm]
\pi^{(\phi)}_{I} & = & \ \dfrac{\partial \mathcal{L}}{\partial \phi'^{I}} & = & - \ e^{2a}\, K_{I J} \,  \phi'^{J} & .
\end{array}
\ee
It may be easily seen that the corresponding Hamilton equations are equivalent to the field equations in \eqref{DW_2nd_order}. 
The associated Hamiltonian $\,\mathcal{H} \,\equiv\, \pi^{(a)} \, a' \, + \, \pi^{(\phi)}_{I} \, \phi'^I \, - \, \mathcal{L}\,$ reads
\be
\mathcal{H}\,=\,\frac{1}{12}\,e^{-2a}\,\left(\pi^{(a)}\right)^2\,-\,\frac{1}{2}\,e^{-2a}\,K^{I J}\,\pi^{(\phi)}_{I}\,\pi^{(\phi)}_{J}\,-\,3\,q_0\,e^{2a}\,+\, 
e^{4a}\,V \ . \nonumber
\ee
Once on-shell, this is an identically vanishing function due to the equation of motion for $a$. 

The HJ formulation of our problem is then constructed by introducing the following \emph{Hamilton's principal function} (HPF) 
\be
S_{\textrm{HPF}} \ \equiv \  F(a,\phi) \ - \ \Psi \ \zeta \ ,
\ee
where the generating functional needs to satisfy the following HJ differential constraint
\be
\label{HJ_non_sep}
0 \, = \, \frac{1}{12} \, e^{-2a} \, \left( \frac{\partial F}{\partial a} \right)^{2} \, - \, 
\frac{1}{2} \, e^{-2a} \, K^{I J} \, \frac{\partial F}{\partial \phi^I} \, \frac{\partial F}{\partial \phi^J} \, - \, 
3 \, q_0 \, e^{2a} \, + \, e^{4a} \, V \ ,
\ee
which represents a PDE for the unknown function $F$ of the $N+1$ variables $\left(a,\,\phi^I\right)$.
It is worth stressing that determining a solution to in \eqref{HJ_non_sep} is generically extremely complicated when $N\,>\,1$. 
However, once in possess of a solution, one has access to the complete dynamical information concerning our second-order problem in \eqref{DW_2nd_order}
through solving the following set of \emph{first-order} flow equations
\be\label{HJ_q0}
\left\{
\begin{array}{lclc}
\dfrac{\partial F }{\partial a} &=& \ \pi^{(a)} & , \\[3mm]
\dfrac{\partial F }{\partial \phi^I} &=&  \pi^{(\phi)}_{I} & .
\end{array}\right.
\ee
For non-zero  values of $q_0$, this equation does \emph{not} admit any solutions obeying a separable \emph{Ansatz}. On the other hand, the equation is easily separable if $q_0 = 0$. 
The fact that the $q_0 \neq 0$ case is more complicated is a direct consequence of the impossibility of decoupling the corresponding equations of motion.

\subsubsection*{The HJ equation for the static $q_0 = 0$ case}

As just discussed above, in the $q_0 = 0$ case, we may write down a factorized \emph{Ansatz} of the form
\begin{equation}
\label{separable_F}
F(a,\phi) \ = \ 2 \, e^{3 a} \, f(\phi) \ ,
\end{equation}
and hence the HJ first-order flow equations \eqref{HJ_q0} take the following simplified form
\be\label{HJ_q0=0}
\left\{
\begin{array}{lclc}
f &=& \  a' \, e^{-a} & , \\[2mm]
\dfrac{\partial f}{\partial \phi^I} &=& - \dfrac{1}{2} \, e^{-a} \, K_{I J} \, \phi'^J & .
\end{array}\right.
\ee
The HJ equation then becomes
\begin{equation} \label{HJEqforf}
V \ = \ - \ 3 \, f^2  \, + \, 2 \, K^{I J} \, \frac{\partial f}{\partial \phi^I} \, \frac{\partial f}{\partial \phi^J} \ , 
\end{equation}
which may be interpreted as the existence of a \emph{fake superpotential} associated with the scalar potential $V$.  
By making use of \eqref{HJ_q0=0} and \eqref{HJEqforf}, one can show that the action \eqref{action_1D_appendix} can be written on-shell as a total
 derivative. This may be viewed as a sign that the above static flows in general define extremal (though not necessarily BPS) DW's.
Such walls are ``flat bubbles'' and, since they have a tension that exactly saturates the bound \eqref{CDL_bound}, they do not represent any gravitational decay
channel. 

\subsubsection*{A change of coordinates}
In order to make contact with the static DW equations used in the main body of the paper, we introduce here the following change of coordinates 
\begin{equation}
e^{a(\zeta)} \, d \zeta \  \equiv \ d z \ , 
\end{equation}
which simplifies the HJ flow equations in the static case by realizing a complete decoupling between the warp factor $a$ and the scalars $\phi^I$. 
In this way, the interval covered by $\zeta$, \emph{i.e.} $[0,+\infty)$, corresponds to $z\, \in \,(-\infty,+\infty) $. 
We may now rewrite the HJ static flow equations \eqref{HJ_q0=0} in these coordinates, which read
\be\label{HJ_q0=0_new}
\left\{
\begin{array}{lclc}
f &=& \  a'  & , \\[2mm]
\dfrac{\partial f}{\partial \phi^I} &=& - \dfrac{1}{2} \, K_{I J} \, \phi'^J & ,
\end{array}\right.
\ee
where now primed quantities represent derivatives with respect to $z$ rather than $\zeta$.

\section{Solving the Hamiton Jacobi equation}\label{appendix:solving_PDEs}

The scope of this appendix is that of presenting a perturbative technique for solving the non-linear PDE in (\ref{HJEqforf}) for a generic case with multiple scalar fields labelled by 
$I\,=\,1,\,\cdots,\,N$.
The focus of our study is to find fake superpotentials $f$ defining static flows starting from a critical point of the scalar potential $V(\phi)$.
Let $\phi_{0}$ be such a critical point, \emph{i.e.} a point where $\partial_I V|_{\phi_{0}} = 0 \ , \ \forall \ I $. 

Static flows of the type of (\ref{HJ_q0=0_new}) starting from $\phi_{0}$ are then characterized by fake superpotentials satisfying
\be\notag
\partial_I f|_{\phi_{0}} = 0  \qquad , \,\forall \ I \ .
\ee

The method that we propose for solving the PDE in (\ref{HJEqforf}) is based on a generalization of the perturbative method sketched in section~\ref{sec:DW} for the case of a single scalar field.
Let us therefore consider a hypothetical solution $f$ written as an expansion in powers of $\Phi^I \,\equiv\, \phi^I - \phi^I_0$ around the critical point $\phi^I_0$. 
We need then to expand the potential $V$ as well in a similar fashion. 
\begin{eqnarray}
V &=& V^{(0)} \ + \ \frac{1}{2} \ V^{(2)}_{IJ} \  \Phi^I \ \Phi^J \ + \ \frac{1}{3!} \ V^{(3)}_{IJK} \  \Phi^I  \ \Phi^J \ \Phi^K \ + \ ... 
\\
f &=& f^{(0)} \ + \ \frac{1}{2} \ f^{(2)}_{IJ} \  \Phi^I \ \Phi^J \ + \ \frac{1}{3!} \  f^{(3)}_{IJK} \  \Phi^I  \ \Phi^J \ \Phi^K \ + \ ... 
\end{eqnarray}

Now, just as explained in section~\ref{sec:DW}, the sets of equations determining the values of the derivatives of $f$ at $\phi_{0}$
decouple from each other order by order upon imposing $\partial_I f|_{\phi_{0}} = 0$.
Equating term by term and in consideration of the previous constraints, one finds
\begin{eqnarray}
f^{(0)} &=& \sqrt{- \frac{1}{3} \ V^{(0)}} \ , \label{f0ord}
\\
f^{(2)} &=& \frac{3}{4} \ f^{(0)} \ K \ + \ \frac{1}{2} \ K^{1/2} \ \sqrt{- \ \frac{3}{4} \ V^{(0)} \ + \ K^{-1/2} \ V^{(2)} \ K^{-1/2} } \ K^{1/2} \ ,\label{f2ord}
\\
 & \vdots & \notag
\end{eqnarray}
where $K$ denotes the kinetic metric for the scalars. Since $f^{(0)}$ purely appears quadratically in the 0$^{\textrm{th}}$ order expansion of (\ref{HJEqforf}), one may pick $f^{(0)}$ to be positive by convention 
without any loss of generality.
The square root in \eqref{f2ord}, instead is meant as a \emph{multivalued} object. The sign choices thereof are discussed in the below. Furthermore, in analogy with the case of a single scalar,
all the other algebraic equations for $f^{(k)}$, with $k\,>\,2$ have degree one and hence admit a unique solution.
As a consequence, there appears to be a discrete set of local branches starting from $\phi_{0}$ and these may be labelled by specifying the eigenvalues of $f^{(2)}_{IJ}$. 

In order to consider all the distinct sign choices for the solutions for $f^{(2)}$ in (\ref{f2ord}), it is enough to consider: 
\begin{itemize}
\item ``$K^{1/2}$'' defined as the unique positive-semidefinite square root of $K$,
\item all the $2^N$ square roots of $\left(- \ \frac{3}{4} \ V^{(0)} \ + \ K^{-1/2} \ V^{(2)} \ K^{-1/2} \right)$.
\end{itemize}
So, after fixing $f^{(0)}$ by following the previous prescription, then each of the $2^N$ solutions for $f^{(2)}$ is fixed and well-defined. Note that a simple way of understanding the arising of the 
$2^N$ local branches can be that of observing that $f^{(2)}$ is symmetric and hence diagonalizable. Therefore, the 2$^{\textrm{nd}}$ order piece of (\ref{HJEqforf}) can be seen as a rank-$N$ system of 
second-degree equations for its eigenvalues. Such a system has $2^N$ real solutions whenever $\phi_{0}$ satisfies the BF bound. It may be worth noticing that the aforementioned solutions need not 
realize all possible sign combinations, as one can already see from the $N\,=\,1$ example solved explicitly in \eqref{Nilsson_branch_SUSY}. This implies that the actual values of the eigenvalues of
$f^{(2)}$ are needed in order to unambiguously specify a certain local branch. These can therefore be used as identifying labels.
 
As anticipated earlier, these labels also turn out to completely determine a unique solution of the remaining full system of equations at arbitrary order. In fact, as long as the contraction of 
$\partial_ I f$ with $(k+1)$-th order derivatives goes to zero when 
$\phi^I \rightarrow \phi^I_0$, the equations for $V^{(k)}$ form a linear system for the $f^{(k)}$ with equal number of equations and unknowns, namely $\binom{N+k-1}{k}$, with $k \geq 3$. 
This does not guarantee that the unknowns will be fully determined as degeneracies can still happen and, most importantly, the system of equations will, by definition, fail if there are any non-analytic 
behaviors arising.

We conclude by providing the actual eigenvalues of $f^{(2)}$ labelling the local branches that we chose in the example of the massive type IIA model on $S^{3}\,\times\,S^{3}$:
\be
\begin{array}{lc}
\textrm{Eig}\left[f^{(2)}(\phi_{1})\right] \ \approx \ \left\{2.11941,\, 1.32674,\, -1.27404,\, -0.52564,\, 0.374264,\, 0\right\} & , \\[2mm]
\textrm{Eig}\left[f^{(2)}(\phi_{3})\right] \ \approx \ \left\{3.48468,\, 3.48468, \,2.43612,\, 2.43612,\, 0.638167,\, 0.638167\right\} & , \\[2mm]
\textrm{Eig}\left[f^{(2)}(\phi_{4})\right] \ \approx \ \left\{4.05124,\, 2.87323, \,2.0047, \,1.65115, \,0.126329, \,0.107406\right\} & , 
\end{array}
\ee
where the labels ``1'', ``3'' \& ``4'' refer to the stable critical points in table~\ref{table:AdS4}, in order to select the local branches defining the globally bounding functions plotted in 
figure~\ref{fig:Stability_6Scalars}, whereas the static DW connecting 1 \& 4 has
\be
\begin{array}{lc}
\textrm{Eig}\left[f^{(2)}(\phi_{1})\right] \ \approx \ \left\{-1.28369,\, -0.888838, \,-0.537439,\, 0.182469,\, -0.173765,\, 0\right\} & , \\[2mm]
\textrm{Eig}\left[f^{(2)}(\phi_{4})\right] \ \approx \ \left\{-1.76277,\, -0.946752,\, -0.700734,\, -0.0337237,\, 0,\, 0\right\} & .
\end{array}
\ee
%

%
%

\small

\bibliography{references}
\bibliographystyle{utphys}

\end{document}